%% file: paper.tex
\documentclass[10pt,journal,compsoc]{IEEEtran}
\IEEEoverridecommandlockouts
\usepackage{cite}
\usepackage{amsmath,amssymb,amsfonts}
\usepackage{algorithmic}
\usepackage{graphicx}
\usepackage{textcomp}
\usepackage{xcolor}

\usepackage{comment}
\usepackage[implicit=false,hidelinks]{hyperref}
\usepackage{booktabs}
\usepackage{float}
\usepackage{cite}
\usepackage{amsmath,amssymb,amsfonts}
\usepackage{algorithmic}
\usepackage{graphicx}
\usepackage{textcomp}
\usepackage{color}
\usepackage{graphicx}
\usepackage{cite}
\usepackage{xspace}
\usepackage{color}
\usepackage{amsmath}
\usepackage{amssymb}
\usepackage{amsfonts,dsfont}
\usepackage{caption}
\usepackage{verbatim}
\usepackage{algorithm}
\usepackage{algorithmic}
\usepackage{multirow}
\usepackage{adjustbox}
\usepackage{soul}
\usepackage{subfig}

\usepackage{diagbox}
\usepackage{float}
\usepackage{amsthm}
\usepackage{booktabs}
\usepackage{caption}
\usepackage{url}
\usepackage{float}
\usepackage{stfloats}

\newtheorem{theorem}{{\bf Theorem}}
\newtheorem{lemma}{{\bf Lemma}}
\newtheorem{definition}{{\bf Definition}}
\newtheorem{property}{{\bf Property}}

\def\BibTeX{{\rm B\kern-.05em{\sc i\kern-.025em b}\kern-.08em
    T\kern-.1667em\lower.7ex\hbox{E}\kern-.125emX}}

\begin{document}

\title{ Deterministic Data Distribution for Efficient Recovery in Erasure-Coded Storage Systems
}

\author{\IEEEauthorblockN{Liangliang~Xu, Min~Lyu, Zhipeng~Li, Yongkun~Li, Yinlong~Xu}
\IEEEcompsocitemizethanks{\IEEEcompsocthanksitem {Liangliang~Xu, Min~Lyu, Zhipeng~Li, Yongkun~Li and Yinlong~Xu
are with the School of Computer
Science and Technology, University of Science and Technology of China,
Hefei 230026, China (e-mail: llxu@mail.ustc.edu.cn; lvmin05@ustc.edu.cn; lizhip@mail.ustc.edu.cn;
ykli@ustc.edu.cn; ylxu@ustc.edu.cn).
}}
}

\IEEEtitleabstractindextext{
\begin{abstract}
Due to individual unreliable commodity components,
failures are common in large-scale distributed storage systems.
Erasure codes are widely deployed in practical storage systems to provide fault tolerance with low storage overhead. However, random data distribution (RDD), commonly used in erasure-coded storage systems, induces   heavy cross-rack traffic, load imbalance, and random access, which adversely affects failure recovery.
In this paper, with orthogonal arrays,
we define a Deterministic Data Distribution ($D^3$) to uniformly distribute data/parity blocks among nodes,    and propose
an efficient failure recovery approach based on $D^3$, which   minimizes the cross-rack repair traffic against a single node failure. Thanks to the uniformity of $D^3$, the proposed recovery approach   balances the repair traffic not only among nodes  within a rack but also
among racks. We implement $D^3$ over Reed-Solomon codes and Locally Repairable Codes in Hadoop Distributed File System (HDFS) with a cluster of 28 machines. Compared with RDD, our experiments show that $D^3$ significantly speeds up the failure recovery up to 2.49 times for  RS codes and 1.38 times for LRCs. Moreover, $D^3$ supports front-end applications better than RDD in both of normal and recovery states.
\end{abstract}
}

\maketitle
\vspace{-5pt}
\input{sections/1-introduction}

\vspace{-5pt}
\input{sections/2-background}

\vspace{-5pt}
\input{sections/3-motivation-example}

\vspace{-5pt}
\input{sections/4-layout}
\vspace{-5pt}
\input{sections/5-recovery}

\vspace{-5pt}
\input{sections/6-evaluation}

\vspace{-5pt}
\input{sections/7-related-work}

\vspace{-5pt}
\input{sections/8-conclusion}

\vspace{-5pt}

\vspace{-5pt}
\section*{Acknowledgments}
This work was supported by the National Key R\&D Program of China under Grant No. 2018YFB1003204, and National Nature Science Foundation of China under Grant No. 61832011 and 61772486. A preliminary version \cite{s3} of this paper was presented at 2019 IEEE International Parallel and Distributed Processing Symposium (IPDPS'19). In this journal version, we extend normal data placement and recovery algorithmic of $D^3$ to Locally Repairable Codes \cite{b18, b7, s7}, provide efficient strategy to maintain the original $D^3$ data layout after recovery, and conduct more experimental evaluation.

\vspace{-10pt}
\bibliographystyle{IEEEtran}
\bibliography{bibfile}

\input{sections/9-appendix}

\end{document}

%% file: sections/1-introduction.tex
\section{Introduction} \label{sec:introdution}
\IEEEPARstart{A} large-scale Distributed Storage System (DSS) typically consists of many individual unreliable commodity components, which induces
frequent component failures \cite{b1, b2}.
In order to guarantee high reliability and availability against component failures,
a common approach is to store data with redundancy.
\emph{Replication} and \emph{erasure coding} are two common approaches to provide fault tolerance.

Replication provides the simplest form of redundancy by storing multiple copies of   data. For example, Google File System (GFS) \cite{b5} and Hadoop Distributed File System (HDFS) \cite{b6} store three copies of each data block by default to tolerate double component failures. Replication is easy to  deploy and recover against failures. However, the storage overhead is too high, say  3$\times$  for triplicate.

As an alternative, erasure codes achieve the same fault tolerance as replication with much lower storage overhead and are commonly deployed in modern DSSes \cite{b7, b8, b10, b2}.
For example,  Google's ColossusFS (the successor to GFS) and Facebook's HDFS-RAID \cite{b10}  deploy Reed-Solomon (RS) codes \cite{b9}, which reduce storage redundancy to 1.5$\times$ and 1.4$\times$ respectively, compared to 3$\times$  in traditional triplicate.
But with erasure codes,
recovering a failed block needs to retrieve multiple available blocks, which induces high repair cost.
For example, in a Facebook's DSS storing multiple petabytes of RS coded data,
a median of 180 terabytes of repair traffic is generated per day \cite{b2}.
Though erasure codes improve storage efficiency,
they significantly increase disk and network traffic for failure recovery.

Compared to RS codes, Locally Repairable Codes (LRCs) are a class of efficiently repairable codes and offer higher reliability with additional local parity blocks\cite{b18, b7, s7}, by which  LRCs  significantly reduce occupation of bandwidth in failure recovery.
Many IT infrastructure companies begin to consider deploying  wide stripes to provide competitive reliability, low storage overhead and high levels of resiliency. In such a  scene\cite{s1}, one file divided into  many data blocks, say 150, and encoded to a small number of parity blocks, say 4,
 the reconstruction bandwidth cost of RS coded stripes  is insufferable, while LRCs  can save lots of network bandwidth, and thus becomes a good choice for the deployment of wide stripes.

Fast recovery  is one key goal for DSSes to avoid secondary failures or irreparable failures.
In addition,  the reconstruction tasks have to compete  with online applications\cite{s6,s4}, e.g. MapReduce tasks and data queries, for system resources,
so in failure recovery, low interference on front-end users' requests is another key goal. Achieving the two goals concurrently in DSSes is interesting and challenging.

A large-scale DSS is deployed over a number of physically independent storage nodes, which are organized into multiple racks, each rack consisting of multiple nodes, and usually adopts random block placement \cite{b8, b2, b18, b20, b21} to achieve load balance among different nodes and different racks.
To maximize the data availability in DSSes deployed with erasure codes,
different blocks of an erasure-coded stripe are stored in nodes of different racks, one block per rack \cite{b1, b28, b7, b8, b18}.
This ``one block per rack" block placement makes a system tolerate the same number of node failures and rack failures.

However, random block placement degrades failure recovery due to two reasons. (1) It can not achieve reconstruction read/write load balance well within short ranges of successive stripes. Such ``local load balancing" is crucial for repair performance \cite{b32}. (2)  Failure recovery based on random block placement induces heavy random access load to retrieve surviving blocks, which will slow down the recovery process.
Furthermore, the ``one block per rack" block placement inevitably makes the repair of any failed block retrieve available blocks from other racks, and as a result, triggers a large amount of cross-rack traffic.
In DSSes, the inner-rack bandwidth is considered to be sufficient, but the cross-rack bandwidth is constrained.
Typically, the available cross-rack bandwidth per node is only 1/20 to 1/5 of the inner-rack bandwidth \cite{b14, e2}.
Thus, cross-rack bandwidth is often considered to be a scarce resource \cite{b16, b17}.
Too heavy cross-rack traffic unavoidably delays the recovery process.

In this paper, we propose a Deterministic Data Distribution ($D^3$) scheme for the data placement in erasure-coded DSSes.
$D^3$ relaxes the constraint of ``one block per rack" and places multiple blocks of the same stripe to different nodes within the same rack to reduce the cross-rack traffic for single-node failure recovery.
Because in DSSes, more than 90\% of component failures are single-node failures,
while double rack failures rarely happen \cite{b1, b31, b8},
$D^3$ still reaches high reliability.
$D^3$ also achieves ``local load balancing" by deterministically distributing data and parity blocks to significantly speed up the failure recovery.
Thanks to orthogonal arrays, $D^3$ ensures read/write load balance even when the storage systems are in different states, such as normal mode and failure recovery, and thus  supports  front-end applications better and has
low interference on the users' requests in recovery.
In summary, our contributions can be summarized as follows.

\begin{itemize}
\item We design the data layout of $D^3$ via orthogonal arrays to
    deterministically distribute blocks to all nodes in the storage systems, and  theoretically prove that $D^3$  provides  uniform data layout for RS codes and LRCs in normal state.

\item Based on $D^3$, we present a recovery algorithm which minimizes the cross-rack repair traffic and achieves load balance of recovery traffic against single node failure. So $D^3$  further guarantees high-quality services for front-end applications.

\item We implement $D^3$ in \textit{HDFS Erasure Coding} \cite{b28}
    shipped with HDFS in Apache Hadoop 3.1.x, and conduct experiments
    with a cluster of 28 machines with different architectures to evaluate its performance.
    Results show that $D^3$ speeds up the failure recovery process  up to 2.49 times for  RS codes and 1.38 times for LRCs compared with random data distribution.

\end{itemize}

The rest of this paper is organized as follows.
We present some background in Section \ref{sec:background}. We motivate this work and
present the main idea of $D^3$ via an example in Section \ref{sec:motivation_example}.
The data layout of $D^3$ and the single-node failure recovery algorithm are presented in  Section \ref{sec:D3}
and Section \ref{sec:recovery} respectively.
The experimental results are presented in Section \ref{sec:evaluation}.
Section \ref{sec:related_work} reviews the related works,
and finally Section \ref{sec:conclusion} concludes this paper.
In order to follow the main logic flow of this paper more easily,
we put the proofs of the lemmas and theorems for $D^3$ in the Appendix.

%% file: sections/2-background.tex
\section{background} \label{sec:background}
In this section, we first introduce the typical architecture of DSSes,
then review two popular categories of erasure codes, Reed-Solomon ($RS$) codes \cite{b9} and
Local repairable codes ($LRCs$) \cite{b18, b7, s7},
and finally introduce orthogonal arrays \cite{b25},
the foundation to define $D^3$.

\vspace{-5pt}
\subsection{Distributed Storage System Architecture}
\begin{figure}
\setlength{\abovecaptionskip}{5pt}
\setlength{\belowcaptionskip}{-15pt}
    \centering
    \includegraphics[width=0.75\linewidth]{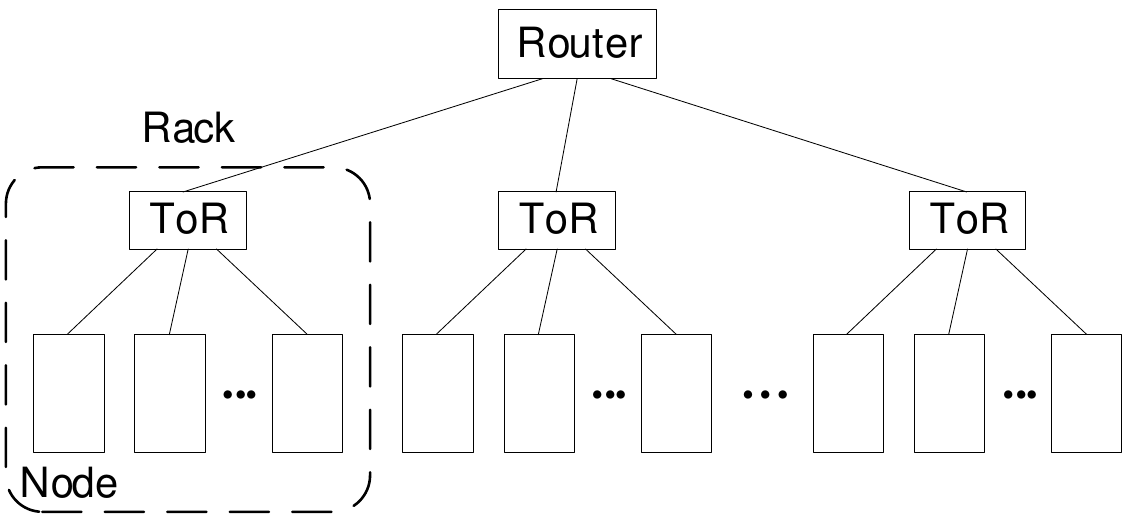}

    \caption{The typical architecture of distributed storage systems.}
    \label{fig:architecture}

\end{figure}

As shown in Fig. \ref{fig:architecture}, a typical distributed storage system (DSS), such as GFS \cite{b5}, HDFS \cite{b6}, Azure \cite{b7}, and Ceph \cite{b20}, consists of some racks, each rack containing multiple nodes (or storage servers). All nodes within the same rack are connected by a top-of-rack (ToR) switch, while all racks are connected by a central router.
The available cross-rack bandwidth suffers from severe competition
because many read or write operations (e.g., degraded read, shuffle/join traffic of computing jobs) access data across different racks \cite{b16}.
As a result, cross-rack bandwidth is considered to be a more scarce resource than inner-rack bandwidth.
So we think that cross-rack data traffic is crucial to fast failure recovery in DSSes.
To store and retrieve a large amount of data,
most DSSes use append-only write and store files as a collection of \emph{blocks} of fixed-size, which form the basic read/write data units.
Such DSSes include GFS \cite{b5}, HDFS \cite{b6}, and Azure \cite{b7}.

\vspace{-5pt}
\subsection{Reed-Solomon Codes}

Given two positive integers, $k$ and $m$,
a $(k, m)$ code encodes $k$ data blocks into $m$ additional \emph{parity} blocks,
such that any one of the $k + m$ blocks can be reconstructed from any other $k$ ones.
All the $k + m$ data/parity blocks form a \emph{stripe} and
$k + m$ is called \emph{stripe size}.
Thus, a $(k, m)$ code tolerates any $m$ blocks being lost in a stripe,
and it achieves the so-called \emph{maximum distance separable} (MDS) property \cite{b26}.
Note that our $D^3$ can be used to any MDS codes.
Since RS codes \cite{b9} are the most popular MDS   codes that are widely deployed in real DSSes \cite{b1, b2, b10}, we present $D^3$ over RS codes and compare its performance with traditional random distribution of blocks in this paper.
Furthermore, RS codes  satisfy the property of \emph{linearity} \cite{b26}.
That is, given a ($k$, $m$)-RS code, any block $B'$ is the linear combination of any other $k$ different blocks $B_0, B_1, \cdots, B_{k-1}$ in the same stripe.
We can get $B'$ by
$B' = \sum_{i=0}^{k-1}c_iB_i$,
where the $c_i$'s ($0 \leq i \leq k-1$) are the decoding coefficients specified by the given RS code.
Note that additions and multiplications here are performed in a finite field.

\vspace{-5pt}
\subsection{Locally Repairable Codes}
Local repairable codes ($LRCs$) are a representative family of non-MDS  codes deployed in Microsoft Azure and Facebook \cite{b18, b7}. For maximum rack-level fault tolerance, each block of one LRC stripe is usually placed in a different rack.
A $(k,l,g)$-LRC \cite{b18, s7, b7} stripe  divides  $k$ data blocks into $l$ local groups, and maintains a parity block (called a {\it local parity block}) for each local group, and meanwhile derives $g$ parity blocks (called {\it global parity blocks}) from the $k$ data blocks similarly with MDS codes. Refer to Fig. \ref{fig:lrc_stripe} for an example. In general, LRCs have the following properties:
\begin{itemize}
  \item Arbitrary $g+1$ node failures can be recovered, and up to $g+l$ failures can also be recovered   if they are information-theoretically decodable \cite{b7}.
  \item Single data block/local parity block can be reconstructed by $\frac{k}{l}$ blocks within its local group.
  \item Global parity block can be reconstructed by other parity blocks.
\end{itemize}
LRC brings competitive properties, and introduces   more types of blocks and different reconstruction flows,
which will bring great challenges in designs of data layout and recovery algorithm to achieve load balance in both of normal and recovery states.

\vspace{-5pt}
\subsection{Orthogonal Array}


Orthogonal array \cite{b25} is a class of combinatorial designs,
which are widely used in the design of experimental setting, coding theory, cryptography, and software testing.
In particular, orthogonal arrays are used to define the data layout of $D^3$ in this paper.


\begin{definition}
    \label{def:oa}
    \cite{s5}
    An orthogonal array OA$(n, k)$ is an $n^2 \times k$ array, $\cal{A}$,
    with entries from a set $X$ of cardinality $n$ such that, within any two columns of $\cal{A}$, every ordered pair of symbols from $X$ occurs in exactly one row of $\cal{A}$.
\end{definition}


From  Definition \ref{def:oa}, we can easily obtain the following  properties of orthogonal arrays, which guarantee the uniform
data distribution for $D^3$, either for normal or single-failure recovery accesses.
\begin{property}
\label{def:pro1}
Each column of array OA$(n, k)$ contains each of the $n$ entries equally often (say $n$ times).
\end{property}
\begin{property}
\label{def:pro2}
Given $ x \in X$ in the $i$-th column,  each of the $n-1$ pairs  $(x, y)$  appears equally often in  the $i$-th and  $j$-th columns (say once), where $y \in X - \{x\}$.
\end{property}

For example, Fig. \ref{fig:rack_layout}(d) shows an orthogonal array OA$(5, 4)$ of the entry set $X = \{0, 1, 2, 3, 4\}$, which is a $25 \times 4$ matrix.
Each entry of $X$ appears 5 times in any column and each ordered pair $(i, j)$ with $i, j \in X$ appears exactly once
in any two columns.  Specifically, for each entry, say $0$, in the 3-rd column, the pair consisting of $0$ and every other entry in the 4-th column appears exactly once, marked with yellow rectangles.

\begin{theorem}
    \label{theorem:oa} \cite{s5}
     Let $n=p_1^{e_1} \cdots p_v^{e_v}$ be the factorization of $n$ into primes and let $k = \min\{p_i^{e_i}+1: i=1,  \cdots, v \}$.
    Then there exists an OA$(n, k)$.
\end{theorem}

Given an integer $n$, Theorem \ref{theorem:oa} tells an integer $k$ for the existence of an OA$(n, k)$.
The construction for such an OA$(n, k)$ is proposed in \cite{b25}. With such construction, there are at least $k-1$ columns being identical in the first $n$ rows, which will be used in the design of $D^3$.
For example, Fig. \ref{fig:rack_layout}(d) shows an OA$(5,4)$,
where all columns are identical in the first five rows.

%% file: sections/3-motivation-example.tex

\section{motivations and an example} \label{sec:motivation_example}
In this section, we first present
our goals for
deterministically distributing blocks in DSSes
based on erasure codes to motivate our work, and then explain the main idea of $D^3$ via an example.
We summarize the major notations used in this paper
in Table \ref{tab:D3_notation}.


\begin{table}[htbp]
\setlength{\abovecaptionskip}{5pt}
\setlength{\belowcaptionskip}{-5pt}
    \centering

    \caption{Major notations used in this paper.}


    \begin{tabular}{p{1.5cm}p{6.5cm}}

        \toprule[1pt]

        \textbf{Notation} & \textbf{Description}  \\

        \midrule[0.5pt]

        $r$ & Number of racks in the system \\

        $n$ & Number of nodes in a rack \\

        $R_i$ & The $i$-th rack in the system \\

        $N_{i,j}$ & The $j$-th node in the $i$-th rack \\

        $k$ & Number of data blocks in a stripe \\

        $m$ & Number of parity blocks in a stripe \\

        $len$ & Stripe size \\

        $N_g$ & Number of groups within a stripe, $N_g = \lceil len/m \rceil$ \\

        $S_i$ & The $i$-th stripe \\

        $d_{i,j}$ & The $j$-th data block in stripe $S_i$ \\

        $p_{i,j}$ & The $j$-th parity block in stripe $S_i$ \\

        $\cal{A}$ & An OA$(n, N_g )$ for load balance within a rack \\

        ${\cal{A'}}$ & An OA$(r, N_g + 1 )$ for load balance in all racks \\

        $G^i_j$ & The $j$-th region-group of the $i$-th stripe region \\

        $G^{i*}_j$ & $G^i_j$ with recovered blocks \\

        $H_i$ & New region-group with recovered blocks of the $i$-th stripe region \\

        $l$ & Number of local parity blocks in an LRC stripe \\

        $g$ & Number of global parity blocks in an LRC stripe \\


        \bottomrule[1pt]
    \end{tabular}

    \label{tab:D3_notation}

\end{table}

\begin{figure*}[htb]
\setlength{\abovecaptionskip}{5pt}
\setlength{\belowcaptionskip}{-10pt}
    \centering
    \includegraphics[width=0.9\linewidth]{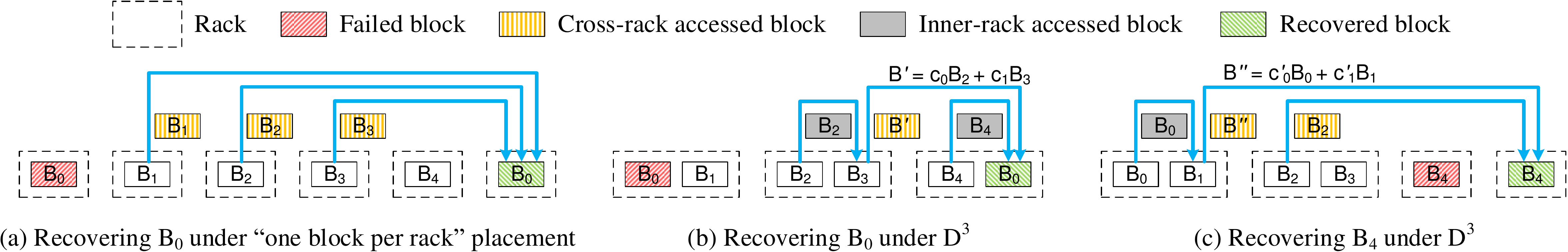}


    \caption{Minimization of cross-rack blocks.}
    \label{fig:stripe_layout}

\end{figure*}

\vspace{-10pt}
\subsection{Motivations} \label{sec:motivation}


To maximize the reliability of DSSes based on erasure codes, blocks are usually distributed such that each rack contains at most one block of the same stripe. Such block placement allows a system to tolerate the same number of node failures and rack failures.
Because each rack contains at most one
block of the same stripe, failure recovery will induce heavy
cross-rack traffic, which prolongs the recovery time. However, rack failures are much rarer than node failures in practical systems, and cross-rack bandwidth is often considered
to be a more scarce resource than inner-rack bandwidth. So
it is viable to tolerate fewer rack failures than node failures
so as to reduce cross-rack repair traffic. Such as in HDFS
and GFS with three replicas for fault tolerance, three copies
of a block are stored in three nodes, where two nodes come
from the same rack, while the other one is in another rack.
So HDFS protects against either double node failures or a
single rack failure. We can distribute blocks of an erasure-coded stripe similarly to HDFS to reduce cross-rack traffic
for failure recovery while keeping high reliability.

Furthermore, DSSes usually distribute blocks to randomly chosen nodes and racks for load balance.
Specifically, in a system deploying a $(k, m)$-RS code or $(k,l,g)$-LRC,
the blocks of the same stripe are stored on different randomly chosen nodes, each node in a separate randomly chosen rack.  
The random distribution achieves uniform distribution of data blocks among all nodes with a huge number of stripes. However, DSSes rebuild lost blocks batch by batch for a long recovery queue due to limited available system resources, such as memory   and CPU, etc.
The random distribution leads to skew distribution of data blocks in a batch with limited stripes, and further induces   random access and load imbalance for rebuilding blocks in a batch. Such random access and ``local load imbalance"
in a batch will slow down the rebuilding process.

This motivates us to design a new scheme for distributing blocks in DSSes based on erasure codes, which takes failure recovery speed into account.
In particular, our design aims for the following three objectives.

\begin{itemize}

    \item \textbf{Objective 1: Uniform Data and Parity Blocks Distribution.}  Each node contains the same amount of data blocks and the same amount of parity blocks
    to achieve load balance.
    Many front-end applications generate intermediate temporary data and also profit from load balance data layout.

    \item \textbf{Objective 2: Minimal Cross-Rack Recovery Traffic.}
    Suppose that there are $t$ nodes in the system.
    Let $s_i$ ($0 \leq i \leq t-1$) be the number of cross-rack accessed blocks
    for recovering the $i$-th node.
    With a data layout against either a single rack failure or $m$ node failures, we aim to minimize $S = \sum_{i=0}^{t-1} s_i$, the total number of cross-rack accessed blocks for recovering all the single-node failures, where $s_i=0$ when the $i$-th node fails, once for a node.  

    \item \textbf{Objective 3: Locally Load-Balanced Recovery Traffic.}
    Under a single node failure, the reconstruction read/write loads are well balanced within a short range of successive stripes among the surviving racks, i.e., the racks not containing the failed node,
    and the read/write/computing loads are balanced among all nodes in the surviving racks, which will significantly speed up the recovery process and be also beneficial to front-end users' services.



\end{itemize}

\vspace{-10pt}
\subsection{$D^3$ Design -- An Example} \label{sec:example}


We now use a simple example of distributing the blocks of a $(3, 2)$-RS code to a storage system with $r=5$ racks and $n=3$ nodes in each rack to show the main idea of $D^3$.
$D^3$ conducts block placement in the following three stages.

\begin{itemize}

    \item \textbf{Minimizing Cross-Rack Accessed Blocks.}
    The lost blocks due to failures can only be recovered with other blocks in the same stripe. So the layout of blocks within a stripe determines the cross-rack traffic for recovery. We firstly design the layout of blocks
    within a stripe to achieve Objective 2.

    \item \textbf{Load Balance within a Rack.}
    Based on the layout of blocks within a stripe, we use an orthogonal array to place blocks of a cluster of stripes so that
    the blocks to be accessed for inner-rack block aggregation and
    the cross-rack accessed blocks for recovery are evenly distributed among nodes within a rack.
    We name a cluster of stripes as a \emph{stripe region}.

    \item \textbf{Load Balance in All Racks.}
    We use another orthogonal array to distribute a group of stripe regions to racks
    to achieve Objective 1 and Objective 3.

\end{itemize}

%
%
%
%
%
%
%
%
%
%
%
%

%


%
%
%

\vspace{-10pt}
\subsubsection{\textbf{Minimizing Cross-Rack Accessed Blocks}}
Let $B_0$, $B_1$, $B_2$, $B_3$, $B_4$ be the five blocks in a stripe of $(3, 2)$-RS code. Suppose we are to reconstruct the failed block $B_0$.
With the ``one block per rack'' placement as shown in Fig. \ref{fig:stripe_layout}(a), we need to access three blocks, say $B_1, B_2, B_3$.
Furthermore,
the recovered block of $B_0$ should be written into another rack different from those storing $B_i$ ($0\leq i \leq 4$).
So totally we access three blocks across racks.

With $D^3$, we first divide the five blocks into three groups as $\{B_0, B_1\}$, $\{B_2, B_3\}$, and $\{B_4\}$, and place each group in a rack with two blocks within a group  in two different nodes for tolerating a single rack failure.
We are still to recover $B_0$.
Because RS codes satisfy the property of linearity,
$B_0$ is a linear combination of $B_2, B_3, B_4$, say
$B_0 = c_0B_2 + c_1B_3 + c_2B_4$,
where the $c_i$'s ($0 \leq i \leq 2$) are the decoding coefficients specified by the given RS code.
As shown in Fig. \ref{fig:stripe_layout}(b), we can first read $B_2$ to the node containing $B_3$, then compute $B' = c_0B_2 + c_1B_3$, read $B'$ and $B_4$ to a new node in the rack containing $B_4$, and at last
recover $B_0$ as $B_0 = B' + c_2B_4$.
So recovering $B_0$ only needs access one block across racks.
Similarly, one block is to be accessed across racks to recover any of $B_1$, $B_2$, and $B_3$.
However, recovering $B_4$ needs to access two blocks across racks,
as shown in Fig. \ref{fig:stripe_layout}(c).
On average, the number of cross-rack accessed blocks for recovering a failed block under $D^3$ is
$(1 \times 4 + 2 \times 1)/5 = 1.2$,
which reaches the minimum with   theoretical guarantee  referred to Lemma \ref{lemma:recovery-stripe-level}.
So the data layout within a stripe achieves Objective 2.

$D^3$ reaches the same reliability against node failures as the ``one block per rack'' placement, but it degrades the reliability against rack failures. Because the probability of double rack failures is negligible, e.g., HDFS also protects against single rack failure as default,
$D^3$ is acceptable against single rack failure.

%

%
%

With grouping  a stripe, we place each group into a rack and distribute  the blocks within a group into different nodes in the corresponding rack.  We use two orthogonal arrays,  ${\cal A}$ and ${\cal A'}$ , for rack-level and node-level distribution, respectively, to achieve load balance.

\subsubsection{\textbf{Load Balance within a Rack}}
\begin{figure}[!h]
\setlength{\abovecaptionskip}{5pt}
\setlength{\belowcaptionskip}{-10pt}
    \centering
    \includegraphics[width=1\linewidth]{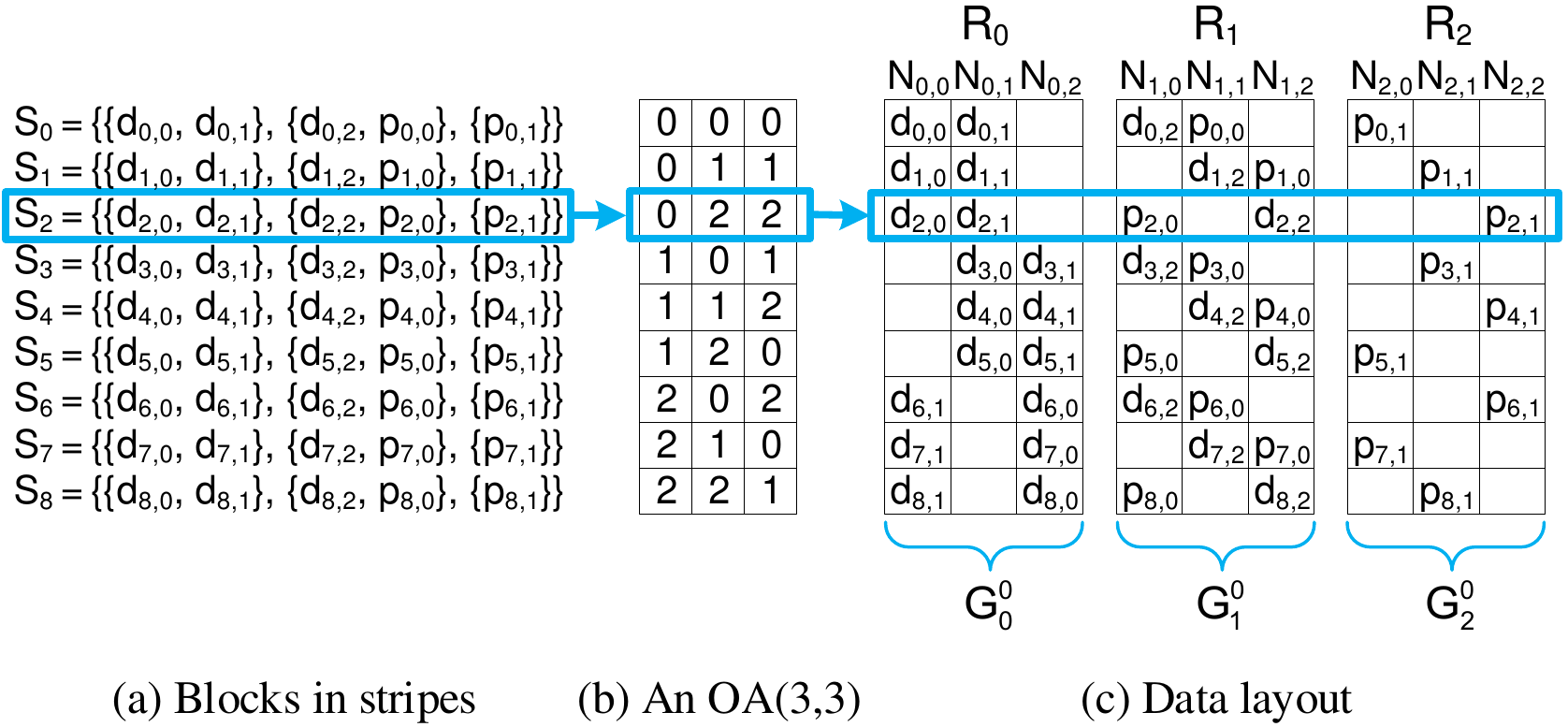}


    \caption{Load balance within a rack.}
    \label{fig:node_layout_1}
\end{figure}

We first display node-level distribution for load balance among all nodes within a rack.
Suppose there are three racks,  $R_0, R_1, R_2$, for placing the three groups of (3, 2)-RS coded stripes, where $R_i$ consists of three nodes, $N_{i,0}, N_{i,1}, N_{i,2}$, for $0 \leq i \leq 2$, as shown in Fig. \ref{fig:node_layout_1}(c).

We select an OA$(3, 3)$, ${\cal A} = (a_{ij})_{9 \times 3}$ (as shown in Fig. \ref{fig:node_layout_1}(b)), where $a_{ij} \in \{0, 1, 2 \}$, as an auxiliary to distribute the blocks.  Each column and each entry of ${\cal A}$ correspond to a rack and a node in the rack respectively. Because there are nine rows in ${\cal A}$, we place a cluster of nine stripes, $S_0, S_2, ..., S_8$, according to ${\cal A}$, as shown in Fig. \ref{fig:node_layout_1}(a).
We partition all blocks in $S_i$ into three groups as above, $g_{i,0}, g_{i,1}, g_{i,2}$, for $0 \leq i \leq 8$, where there are two blocks in each of $g_{i,0}, g_{i,1}$, but only one block in $g_{i,2}$.
According to ${\cal A}$, we place two blocks in $g_{i,j}$ to two nodes $N_{j, a_{ij}}$ and $N_{j,(a_{ij}+1) \bmod 3}$ in rack $R_j$ for $j= 0,1$ respectively, and place the block in $g_{i,2}$ to node $N_{2, a_{i2}}$ in rack $R_2$, as shown in Fig. \ref{fig:node_layout_1}(c).
For example, the third row of ${\cal A}$ is $(0, 2, 2)$.
So we should place blocks $d_{2,0}$ and $d_{2,1}$ in the first group to $N_{0, 0}$ and $N_{0,1}$ in $R_0$ respectively, $d_{2,2}$ and $p_{2,0}$ to $N_{1, 2}$ and $N_{1,0}$ in $R_1$ respectively, and $p_{2,1}$ to $N_{2, 2}$ in $R_2$.
We find that the blocks in a cluster of nine stripes are evenly distributed among the nodes within a rack.
We name a cluster of nine stripes as a \emph{stripe region},
and we divide the $i$-th stripe region into three \emph{region-groups},
$G^i_{0}$, $G^i_{1}$, and $G^i_{2}$,
where $G^i_j$ is in rack $R_j$ for $0 \leq j \leq 2$.

Note that there are two types of group-size in (3, 2)-RS coded stripes, say $2$ of $g_{i,0}$ and $g_{i,1}$ and 1 of $g_{i,2}$. We can distinguish two types of region-groups   with recovered blocks corresponding to whether the recovered blocks are placed in a new rack.

\vspace{-5pt}
\subsubsection{\textbf{Load Balance in All Racks}}

\begin{figure}[!t]
\setlength{\abovecaptionskip}{5pt}
\setlength{\belowcaptionskip}{-10pt}
    \centering
    \includegraphics[width=1\linewidth]{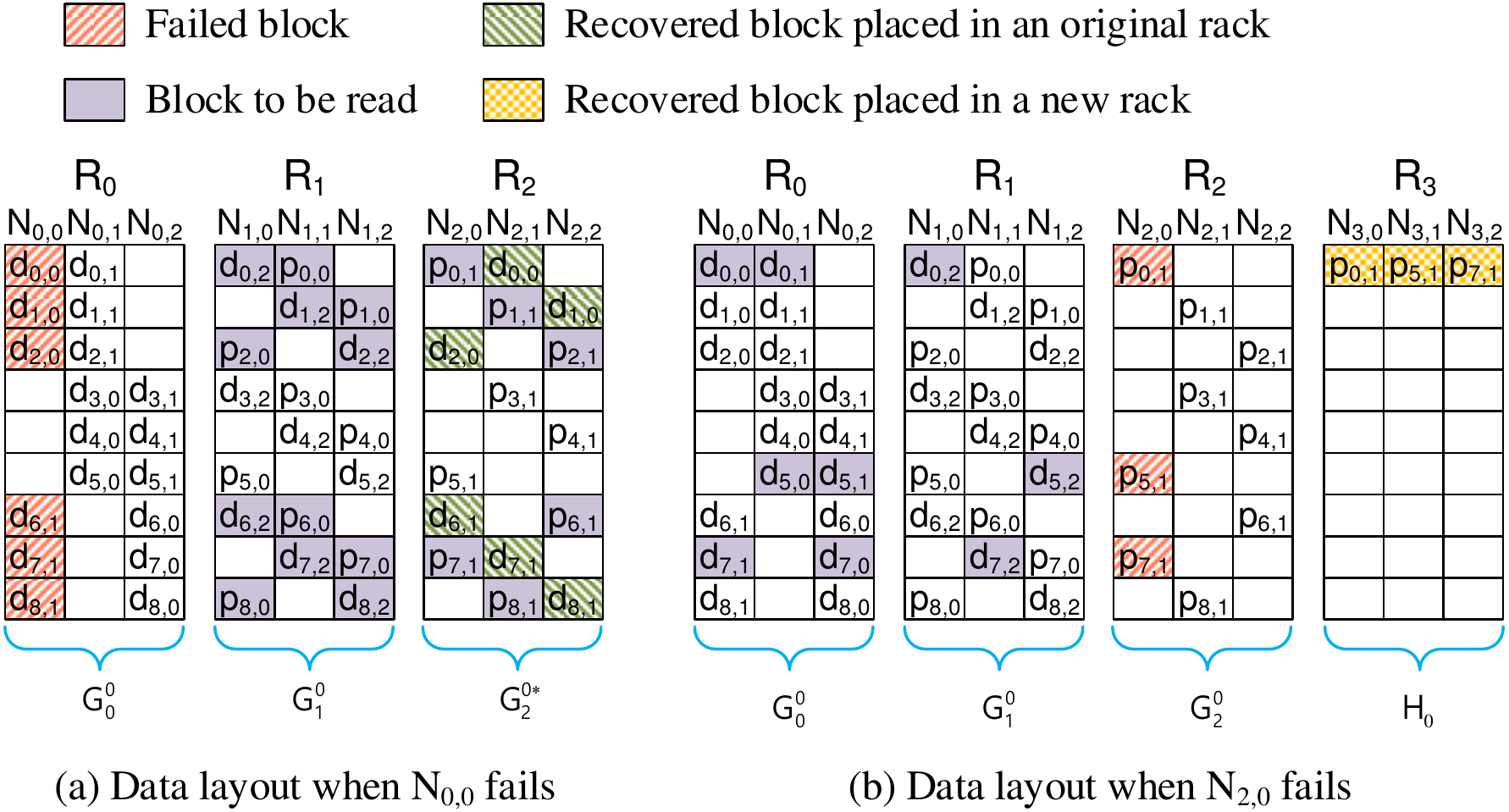}


    \caption{Load balance within a rack when one node fails.}

    \label{fig:node_layout_2}

 \end{figure}

Suppose  node $N_{0,0}$ in rack $R_0$ fails (Fig. \ref{fig:node_layout_2}(a)).
Since $R_2$ contains less blocks than $R_0$,  we store the recovered blocks on  the nodes in $R_2$, $N_{2, (a_{i2}+1) \bmod 3}$, for $i =0,1,2,6,7,8$.
Thanks to Property 2 of orthogonal arrays,
the blocks needed to be accessed in racks $R_1$ and $R_2$  are evenly distributed among the nodes in each rack,
and the recovered blocks are also  evenly distributed among the nodes in $R_2$. The changed  region-group $G^i_2$ is the first type of region-groups  with recovered blocks, denoted as $G^{i*}_2$.

Suppose a node in $R_2$, say $N_{2,0}$, fails.
 Since the recovered blocks need to be placed in a rack different from $R_2$ and the first two racks $R_0$ and $R_1$ contain more blocks than $R_2$, we distribute the  recovered blocks to  three nodes in a new rack $R_3$ evenly in a round-robin way
as shown in Fig. \ref{fig:node_layout_2}(b), which forms another type of  region-group with recovered blocks, denoted as $H_i$. Similarly, the blocks needed to be accessed in racks $R_0$ and $R_1$ are evenly distributed among the nodes in each rack,
and the recovered blocks are also evenly distributed among the nodes in $R_3$.

 \begin{figure*}[!htb]
 \setlength{\abovecaptionskip}{5pt}
\setlength{\belowcaptionskip}{-10pt}
    \centering
    \includegraphics[width=0.9\linewidth]{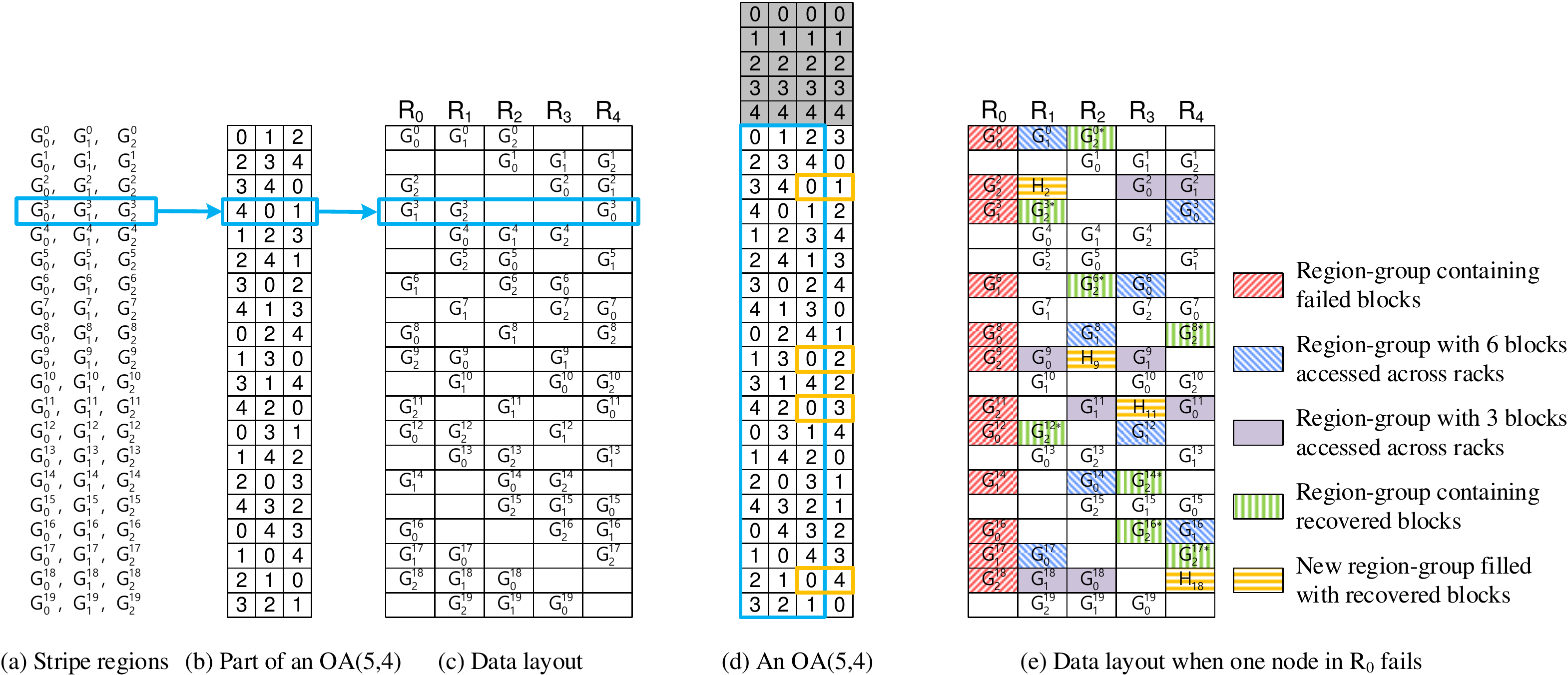}


    \caption{Load balance in all racks.}
    \label{fig:rack_layout}

 \end{figure*}

We have distributed the blocks within a stripe region to reach load balance within a rack.
Now we distribute $r(r-1) = 20$ stripe regions to a system of $r=5$ racks to reach load balance among the nodes in all racks.
We select an OA$(5, 4)$, ${\cal A'} = (a'_{ij})_{25 \times 4}$ as shown in Fig. \ref{fig:rack_layout}(d) to define the data placement of  20 stripe regions, where  $a'_{ij} \in \{0,1,2,3,4\}$ corresponds to rack $R_{a'_{ij}}$.
Note that the  5 columns  in  the first five rows of ${\cal A'}$ are identical. If we use the first five rows of ${\cal A'}$ to define the placement of a stripe region, then all groups in the stripe will be placed in the same rack, which does not tolerate any rack failure. So we only use the sixth to twenty-fifth rows to define the  placement,
denoted as another matrix ${\cal M} = (m_{ij})_{20 \times 4}$, i.e., $m_{ij} = a'_{i+5,j}$.

We diagram the step of placing  stripe regions to reach load balance among all nodes in all racks
in Fig. \ref{fig:rack_layout},
where each row of Fig. \ref{fig:rack_layout}(a) corresponds to a stripe region,
Fig. \ref{fig:rack_layout}(b) shows the first three columns of ${\cal M}$ which define the placement of the region-groups,
and Fig. \ref{fig:rack_layout}(c) shows the placement.
We distribute $G^i_0$, $G^i_1$, $G^i_2$ to racks $R_{m_{i0}}$, $R_{m_{i1}}$, $R_{m_{i2}}$ respectively, for $0 \leq i \leq 19$.
For example, $(m_{30}, m_{31}, m_{32}) = (4, 0, 1)$,
we distribute $G^3_0$, $G^3_1$, $G^3_2$ to racks $R_4$, $R_0$, $R_1$ respectively,
as shown in the fourth row of Fig. \ref{fig:rack_layout}(c).
For any $j$ ($0 \leq j \leq 2$),
the 20 region-groups $G^i_j$'s ($0 \leq i \leq 19$) are evenly distributed among all racks.
Thus, data blocks and parity blocks are evenly distributed among all nodes,
i.e., we achieve Objective 1.

Suppose that one node in rack $R_0$ fails, as shown in Fig. \ref{fig:rack_layout}(e).
There are 12 region-groups in $R_0$ containing a failed block each.
These region-groups in $R_0$ can be categorized into two types according to group size,
$G^i_j$'s for $j = 0, 1$ and $G^i_2$'s.
Because of the even distribution of region-groups, there are four $G^i_j$'s in $R_0$ for each of
$j=0,1$ (see Fig. \ref{fig:rack_layout}(e)), the recovery method for each is similar with Fig. \ref{fig:node_layout_2}(a)
and corresponding region-group with recovered blocks $G_2^i$ become $G_2^{i*}$.
There are also four $G^i_2$'s in $R_0$,
and the each corresponding region-group with recovered blocks $H_i$ (see Fig. \ref{fig:node_layout_2}(b)) is placed into rack $R_{m_{i3}}$.

From Figs. \ref{fig:node_layout_2}(a), \ref{fig:stripe_layout}(b)
and Figs. \ref{fig:node_layout_2}(b), \ref{fig:stripe_layout}(c),
we find that a rack containing a region-group $G^{i*}_2$ or $H_i$ will induce
six cross-rack blocks written into it.
Furthermore,
from Fig. \ref{fig:rack_layout}(e), we find that
both of $G^{i*}_2$'s and $H_i$'s are evenly distributed among the surviving racks.
So the load of cross-rack write for recovering failed blocks is balanced among the surviving racks.
Similarly, the load of cross-rack read for recovering failed blocks is also balanced among the surviving racks
as shown in Fig. \ref{fig:rack_layout}(e).
Since ${\cal A'}$ balances each type of region-groups (e.g., a type of region-groups as $H_i$'s) involved in recovery
among the surviving racks,
the loads of read/write/computing for reconstructing the failed blocks are balanced among all nodes in the surviving racks.
Thus, we achieve Objective 3.

%% file: sections/4-layout.tex
\vspace{-5pt}
\section{The General Design of $D^3$} \label{sec:D3}


In this section, we present the general design of $D^3$.
Suppose we are to deploy  $(k, m)$-RS code to a system composed of $r$ racks with $n$
nodes each, i.e., $n \times r$ nodes in total.
$D^3$ conducts block placement in three stages, data layout of a stripe, load balance within a rack, and load balance in all racks.

\vspace{-5pt}
\subsection{Data Layout of a Stripe} \label{sec:stripe-level}
On one hand, in order to minimize cross-rack repair traffic,
we place multiple blocks of a stripe in a rack
at the expense of reducing rack-level fault tolerance.
On the other hand,
a rack may contain at most $m$ blocks of the same stripe for tolerating a single rack failure.
Thus, $D^3$ divides the $len = k+m$ blocks of a stripe into $N_g = \lceil len/m \rceil$ groups, $g_0, g_1, ..., g_{N_g-1}$, such that there are $Size_{max} = \lceil len/N_g \rceil $ blocks in each of the first $t = len \bmod N_g$ groups $g_0, g_1, ..., g_{t-1}$ and $Size_{min} = \lfloor len/N_g \rfloor $ blocks in each of the remaining $N_g - t$ groups $g_t, g_{t+1}, ..., g_{N_g-1}$.
Specifically, let $B_{0}, B_{1}, \cdots, B_{k+m-1}$ be the $k+m$ blocks in a stripe. $D^3$ allocates $B_{0}, \cdots,$ $B_{Size_{max}-1}$ to $g_0$, $B_{Size_{max}}, \cdots, B_{2 \times Size_{max}-1}$ to $g_1$, ..., $B_{(t-1) \times Size_{max}}, \cdots, B_{t \times Size_{max}-1}$ to $g_{t-1}$, $B_{t \times Size_{max}}, \cdots, B_{t \times Size_{max}+Size_{min}-1}$ to $g_{t}$ and so on. So the allocation of all blocks in a stripe into $N_g$ groups is determined and unique.

Each group will be placed into a separate rack, as shown in Fig. \ref{fig:stripe_layout}(b).
We have the following two lemmas.

\begin{lemma}
    \label{lemma:stripe-level-1}
    With the data layout of $D^3$, there are at most $m$ blocks of the same stripe in a group.
\end{lemma}

\begin{lemma}
    \label{lemma:stripe-level-2}

    Suppose that $len = am + b$, where $a = \lfloor len/m \rfloor$ and $b = len \bmod m$.
    With the data layout of $D^3$, if $0 < b < m-1$, then there are at least two groups, where there are no more than $m-1$ blocks of the same stripe in each group.

\end{lemma}

\vspace{-10pt}
\subsection{Load Balance within a Rack} \label{sec:node-level}

Now we are to place a group of stripes to reach load balance among the $n$ nodes ($n \geq m$) within a rack. Because there are $n$ nodes in each rack, and we divide the blocks of each stripe into $N_g$ groups, we need to distribute a cluster of $n^2$ stripes to the nodes in $N_g$ racks to reach load balance among all nodes within a rack. Denote the $n^2$ stripes as $S_0, S_1, ..., S_{n^2-1}$, where $S_i$ is divided into $N_g$ groups $g_{i,0}, g_{i,1}, ..., g_{i,N_g-1}$ for $0 \leq i \leq n^2-1$.
We select an OA$(n, N_g)$, ${\cal A} = (a_{ij})_{n^2 \times N_g}$, to define the block placement.
The $k$-th block in $g_{i,j}$ is placed to $N_{j,((a_{ij} + k) \bmod n)}$ in rack $R_{j}$ for $0 \leq k \leq Size_{max}$ when there are $Size_{max}$ blocks in $g_{i,j}$; otherwise for $0 \leq k \leq Size_{min}$ when there are $Size_{min}$ blocks in it.
Refer to Fig. \ref{fig:node_layout_1}(c) as an example.
Based on the block placement, we have the following lemma.

\begin{lemma}
    \label{lemma:node-level}
    Within a cluster of $n^2$ stripes, $D^3$ places the same number of data/parity blocks
    to each node in the same rack.
\end{lemma}

\vspace{-10pt}
\subsection{Load Balance in All Racks} \label{sec:rack-level}
Denote a cluster of $n^2$ stripes as a \emph{stripe region}. We now distribute $r(r-1)$ stripe regions to a system of $r$ racks ($ r > N_g $) to reach load balance among all nodes in all racks.
We select an OA$(r, N_g + 1 )$, ${\cal A'}$, with its first $r$ rows of all $N_g + 1$ columns being identical.
We ignore the first $r$ rows of ${\cal A'}$ and denote the submatrix of ${\cal A'}$ without the first $r$ rows as ${\cal M}$, which is used to define block placement of $r(r-1)$ stripe regions.
The last column of ${\cal M}$ is used for failure recovery.

Denote the $i$-th stripe in the $j$-th stripe region as $S^{j}_{i}$, where $S^{j}_{i}$ is divided into $N_g$ groups $g^j_{i,0}, g^j_{i,1}, ..., g^{j}_{i,N_g-1}$ for $0 \leq i \leq n^2-1, 0 \leq j \leq r(r-1)-1$. According to the group division of stripes, we divide the $j$-th stripe region into $N_g$ \textit{region-groups}, $G^j_{0}, G^j_{1}, \cdots, G^j_{N_g-1}$, where $G^j_k$ consists of $g^j_{0, k}, g^j_{1,k}, ..., g^{j}_{n^2-1,k}$ for $0 \leq k \leq N_g-1$.
Suppose ${\cal M} = (m_{jk})_{(r(r-1))\times (N_g+1)}$.
We place the $k$-th region-group of the $j$-th stripe region, $G^j_{k}$, into rack $R_{m_{jk}}$ for $0 \leq j \leq r(r-1)-1, 0 \leq k \leq N_g-1$.
Refer to Fig. \ref{fig:rack_layout}(c) as an example.
The following theorem shows that $D^3$ achieves Objective 1.


\begin{figure}[!tp]
\setlength{\abovecaptionskip}{5pt}
\setlength{\belowcaptionskip}{-10pt}
    \centering
    \includegraphics[width=1\linewidth]{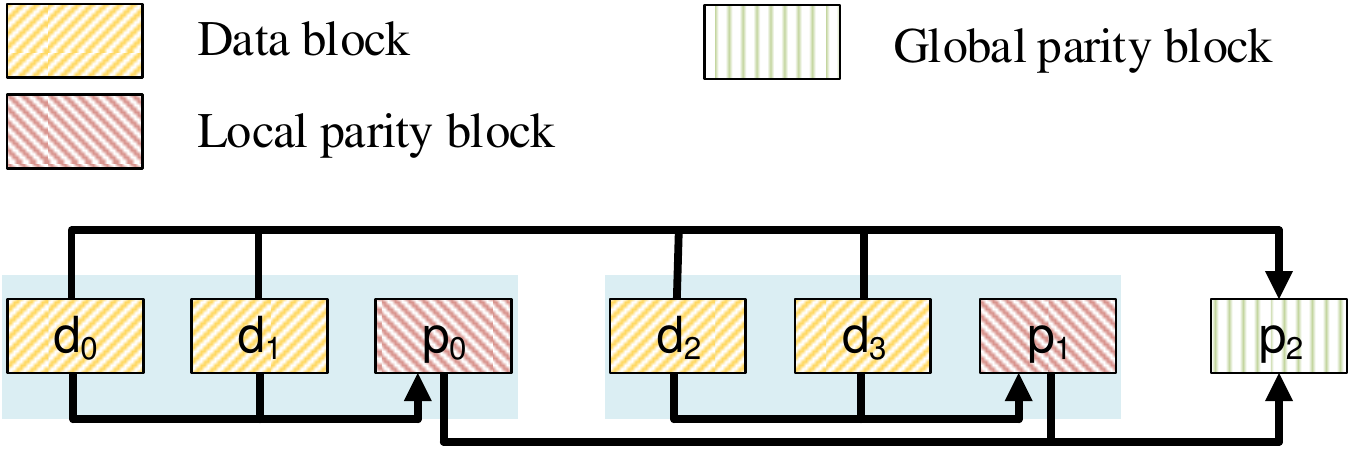}
    \caption{A $(4,2,1)$-LRC stripe. ($k = 4$ data blocks, $l = 2$ local parity blocks and $r = 1$ global parity block.)}
    \label{fig:lrc_stripe}
 \end{figure}

\begin{figure*}[!htbp]
\setlength{\abovecaptionskip}{5pt}
\setlength{\belowcaptionskip}{-10pt}
    \centering
    \includegraphics[width=0.9\linewidth]{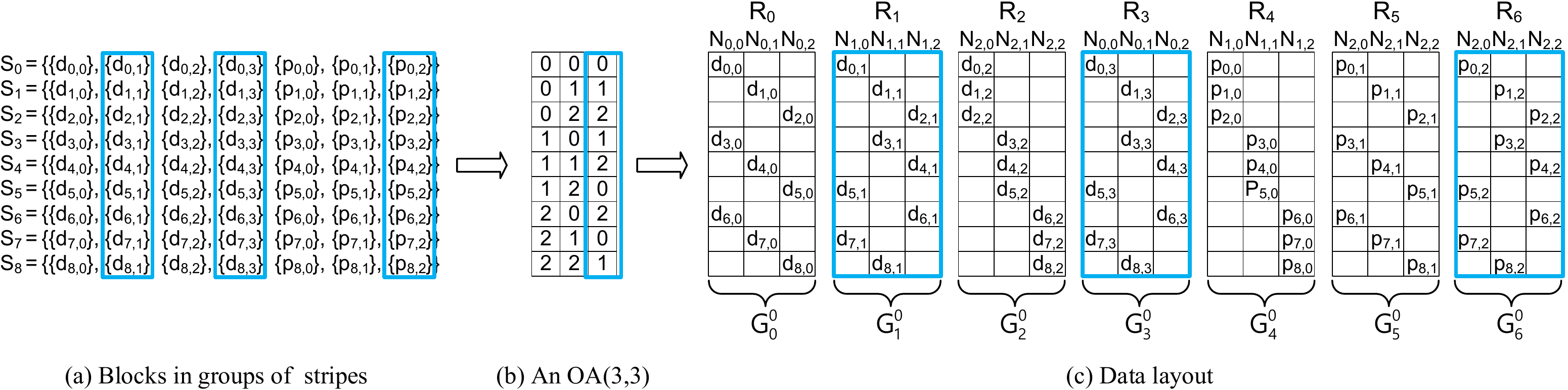}
\caption{ \small Load balance within a rack for $(4,2,1)$-LRC. \{$d_{i,0}, d_{i,1}, p_{i,0}$\} and \{$d_{i,2}, d_{i,3}, p_{i,1}$\} are in two local groups of LRC, \{$p_{i,2}$\} is the global parity block. \{$p_{i,0}, d_{i,2}$\}, \{$d_{i,0}, p_{i,1}$\} and \{$d_{i,1}, d_{i,3}, p_{i,2}$\} use the three columns of OA to address ids of nodes, respectively.}
    \label{fig:lrc_datalayout}
 \end{figure*}

\begin{theorem}
    \label{theorem:rack-level-1}
    Within $r(r-1)$ stripe regions, $D^3$ places the same number of data/parity blocks to
    each node in all racks.
\end{theorem}

The following theorem shows the reliability of $D^3$.

\begin{theorem}
    \label{theorem:rack-level-2}
    $D^3$ tolerates either $m$ node failures or one rack failure.
\end{theorem}

\vspace{-10pt}
\subsection{Extension to Locally Repairable Codes} \label{data layout of lrc}
A $(k,l,g)$-LRC \cite{b18} stripe contains $k$ data blocks, $l$ local groups and $g$ global parity blocks.
Fig. \ref{fig:lrc_stripe} gives an example of $(4,2,1)$-LRC.
The storage overhead and repair traffic of LRCs are between traditional replication and RS  codes.
LRC keeps additional local parity blocks and introduces different reconstructing methods for different types
of failed blocks. So, how to balance repair traffic becomes more difficult.  In view of this, we extend $D^3$ to support LRC to improve repair performance. The block
placement for LRCs contains similar stages with RS
codes except skipping the first stage. The details are shown as follows.

\vspace{-5pt}
\subsubsection{Load balance across nodes}
 In order to satisfy maximum rack-level fault tolerance, we place blocks in a region adhering to the following basic rules: (1) blocks within  a local group placed in different racks, (2) all parity blocks placed in different racks. So, we select an OA$(n,N_g^{lrc})$, ${\cal A} = (a_{ij})_{n^2 \times N_g^{lrc}}$, for
distributing blocks in node level, where $N_g^{lrc} = Max\{\frac{k}{l}+1, l+g\}$.
The following strategy can achieve the above assignments: (1) assign a column for each parity block in a stripe; (2) for each data block in a local group, assign a column different from  the one assigned to the local parity block.
In this way, a small orthogonal array suffices by   one column of ${\cal A}$
specifying the location for multiple blocks instead
of a block within one stripe. Based on the
Property 1 of OA, we achieve load balance for LRCs across nodes whether within a rack or not.

For example in Fig. \ref{fig:lrc_datalayout},
we assign the $(4,2,1)$-LRC stripes to a DSS with OA$(3,3)$ in Fig. \ref{fig:lrc_datalayout}(b).
For illustrating corresponding rules, we use the last column of OA$(3,3)$ as the example.
We first look up the last column of OA$(3,3)$, and then assign data blocks and global parity blocks, saying $d_{i,j}$ and $p_{i,2}$ for $0 \leq i \leq 8$ and $j=1,3$,
to  three racks, say $R_1$, $R_3$ and $R_6$, with each entry specifying the ids of nodes in each rack.
Similarly for other block, we achieve uniform blocks distribution in Fig. \ref{fig:lrc_datalayout}(c).

\vspace{-5pt}
\subsubsection{Load balance across racks}
Now, we will assign $N_g = k + l + g$ region-groups  to $r$ racks and keep rack-level load balance (without loss of generality,  $r>N_g$).
Similarly to Section \ref{sec:rack-level}, we will find an OA$(r,N_g+1)$, ${\cal A'}$,
and denote the submatrix of ${\cal A'}$ without the first $r$ rows as ${\cal M}$,
the first $N_g$ columns of ${\cal M}$ act as an addressing table for
the $k + l + g$
 region-groups of each $(k,l,g)$-LRC stripe region, and the last
column provides the address for the reconstructed blocks
against single node failure.
Then we place $r(r-1)$ regions, each of which consists of $n^2$ stripes, to racks in cluster based on the selected
${\cal M}$ with $r(r-1)$ rows.
The whole step is similar to  the step of load balance in rack level for RS codes in Section \ref{sec:rack-level}.
So based on the properties of OA, we achieve load balance for LRCs in rack level.

From the above designs, we get the following theorem:
\begin{theorem}  \label{theorem:lrc-data-layout}
    $D^3$ places the same number of data blocks, local parity blocks and global parity blocks to each node in all racks for Locally Repairable Codes.
\end{theorem}

\vspace{-5pt}
\subsection{Valid Node Sizes and Rack Sizes of $D^3$}

$D^3$ uses two orthogonal arrays to define its data layout. One is $\cal{A}$, for load balance within a rack, and the other is $\cal {A}'$,
for load balance among nodes in all racks.
From Theorem \ref{theorem:oa}, given an integer $n$,
let $n=p_1^{e_1} \cdots p_v^{e_v}$ and
$s = \min\{p_i^{e_i}+1: i=1,  \cdots, v \}$,
and there is an OA$(n, s)$, where the $s-1$ columns of the first $r$ rows are identical \cite{b25}.
From Definition \ref{def:oa}, for any $s' \leq s$, we can get an OA$(n, s')$ just by taking any $s'$ columns from OA$(n, s)$.

The data layout of $D^3$ is defined by $\cal{A}$ and $\cal {A'}$, where $n$ and $r$ correspond to the number of nodes in a rack and the  number of racks in a DSS respectively. From the construction of orthogonal arrays in \cite{b25},
we know that there are plenty of OA$(n, s)$'s for reasonable $n,s$. So we can obtain $\cal{A}$ and $\cal{A'}$ for practical configurations of DSSes.

%% file: sections/5-recovery.tex
\section{Failure Recovery} \label{sec:recovery}
In this section, we successively present recovery solutions for RS codes
and LRCs based on  $D^3$ against single node failure.
Finally, we present a migration algorithm to maintain the original uniform data layout.

\vspace{-5pt}
\subsection{Recovery for RS codes}
\subsubsection{Recovery Process within a Stripe}

Let $B_0, B_1, \cdots, B_{len-1}$ be the $len=k+m$ blocks in a stripe.
From Section \ref{sec:stripe-level}, the $len$ blocks are divided into $N_g = \lceil len/m \rceil$ groups such that there are $t = len \bmod N_g$ groups, each of which containing $Size_{max} = \lceil len/N_g \rceil$ blocks, and there are $N_g - t$ groups, each of which containing $Size_{min} = \lfloor len/N_g \rfloor$ blocks. Now assume one block of a stripe fails.
Let $\mu$ be the average number of cross-rack accessed blocks for recovering the failed block.
Suppose that $len = am + b$, where $a = \lfloor len/m \rfloor$ and $b = len \bmod m$.
We have the following three cases.

\begin{itemize}

    \item[(1)]
    If $b=0$,
    we have $N_g = a$,
    $t = 0$,
    and $Size_{min} = Size_{max} = m$, i.e., each group contains $m$ blocks.
    We use the $k = (a-1)m$ blocks from the $a-1$ surviving groups
    to reconstruct the failed block as follows.
    In each surviving group,
    the node containing the block with the largest subscript
    reads the other $m-1$ blocks in the same group
    and performs the inner-rack block aggregation,
    and then sends the aggregated block to some node in a new rack,
    where the failed block is reconstructed with the $a-1$ aggregated blocks.
    So we have $\mu = a-1$ in this case.

    \item[(2)]
    If $0 < b < m-1$,
    we have $N_g = a + 1$.
    By Lemma \ref{lemma:stripe-level-2},
    there are at least two groups which contains no more than $m-1$ blocks.
    Suppose that $R_x$ is the rack with largest subscript $x$ which holds a surviving group with $z \leq m-1$ blocks.
    We select $k-z$ blocks with the smallest subscripts
    from the $a-1$ surviving groups excluding the group in $R_x$.
    In each of the surviving groups containing any of these $k-z$ selected blocks,
    the node containing the selected block with the largest subscript
    in this group reads the other selected blocks in the same group
    and performs the inner-rack block aggregation,
    then sends the aggregated block
    to a node $N_y$ in rack $R_x$,
    where $N_y$ does not contain any of the $z$ blocks in $R_x$.
    $N_y$ uses the $a-1$ aggregated blocks and the $z$ blocks in $R_x$ to compute the failed block.
    So we have $\mu = a-1$ in this case.

    \item[(3)]
    Otherwise $b=m-1$,
    then $N_g = a+1$,
    $t = a$, $Size_{max} = m$, $Size_{min} = m-1$. There are $a$ groups, each of them containing $m$ blocks, and one group containing $m-1$ blocks.

        \quad (3.1) If the failed block is in a group containing $m$ blocks,
        then there are $a-1$ surviving groups, each containing $m$ blocks.
        In each surviving group containing $m$ blocks,
        the node containing the block with the largest subscript
        reads the other $m-1$ blocks in the same group
        and performs the inner-rack block aggregation,
        and then sends the aggregated block to node $N_x$,
        where $N_x$ is in the rack storing the group with $m-1$ blocks and
        $N_x$ does not contain any of the $m-1$ blocks in this rack.
        $N_x$ uses the $a-1$ aggregated blocks and the $m-1$ blocks in the local rack to recover the failed block.
        It reads $a-1$ blocks across racks.
        Refer to Fig. \ref{fig:stripe_layout}(b) as an example.

        \quad (3.2) Otherwise, the failed block is in the group containing $m-1$ blocks.
        We select $k = am-1$ blocks with the $k$ smallest subscripts
        from the $a$ surviving groups, i.e., only excluding the block with the largest subscript in the $a$ surviving groups.
        In each surviving group,
        the node containing the selected block with the largest subscript in this group
        reads the other selected blocks in the same group
        and performs the inner-rack block aggregation,
        and then sends the aggregated block to some node $N_x$ in a new rack.
        $N_x$ uses the $a$ aggregated blocks to recover the failed block.
        It reads $a$ blocks across racks.
        Refer to Fig. \ref{fig:stripe_layout}(c) as an example.

        So in case of $b=m-1$, we have $\mu = [(a-1)(k+1) + a(m-1)]/(k+m)$.

\end{itemize}

Furthermore, we have the following lemma.

\begin{lemma}
    \label{lemma:recovery-stripe-level}

    Given a $(k, m)$-RS code,
    suppose that $len = k + m = am + b$, where $a = \lfloor len/m \rfloor$ and $b = len\bmod m$.
    Then, with the data layout of $D^3$,
    the average number of cross-rack accessed blocks for
    recovering a failed block within a stripe is
   \begin{equation}\label{eq:minimum}
    \mu =
    \begin{cases}
    \displaystyle \frac{(a-1)(k+1) + a(m-1)}{k+m} \, , &  \text{if } b=m-1; \\
    a-1 \, , & \text{otherwise},
    \end{cases}
   \end{equation}
    and reaches the minimum average number of cross-rack accessed blocks for recovering a failed block for data layouts against a single rack failure.

\end{lemma}

\vspace{-15pt}
\subsubsection{Recovery Process within a Stripe Region}

Based on the recovery process within a stripe, we place the recovered blocks to nodes as follows.

\begin{itemize}

    \item[(1)]
    Suppose we are to place a recovered block $B$ of stripe $S_i$ to a node in its original rack $R_x$.
    Suppose that the block with the largest subscript
    stored in $R_x$ and belonging to $S_i$ is placed to node $N_{x,j}$.
    Then $D^3$ places $B$
    to node $N_{x,((j + 1) \bmod n)}$ in $R_x$.
    Refer to Fig. \ref{fig:node_layout_2}(a) as an example.

    \item[(2)]
    Otherwise, the recovered blocks are placed into a new rack.
    $D^3$ places the recovered blocks to nodes of the new rack in the round-robin order.
    Refer to Fig. \ref{fig:node_layout_2}(b) as an example.

\end{itemize}

Based on the recovery process within a stripe region,
we have the following lemma.

\begin{lemma}
    \label{lemma:recovery-node-level}

    To recover a failed node, $D^3$ achieves load balance among all nodes within a rack in terms of overheads of read, write, and computing respectively.

\end{lemma}

\vspace{-15pt}
\subsubsection{Recovery Process in $r(r-1)$ Stripe Regions}
The repair loads are balanced among all nodes in surviving racks in $r(r-1)$ stripe regions with $D^3$.
From Section \ref{sec:node-level},
we know that each stripe region is divided into $N_g$ region-groups, $G_{0}, G_{1}, \cdots, G_{N_g-1}$.

\begin{itemize}

    \item[(1)]
    If the recovered blocks of $G^j_{i}$ in the $j$-th stripe region are placed in the rack with $G^j_{x}$, then $G^j_{x}$ changes and the changed one is denoted as $G^{j*}_{x}$, referring to Fig. \ref{fig:rack_layout}(e) as an example.

    \item[(2)]
    Otherwise, the recovered blocks of $G^j_{i}$ are placed in a new rack, forming a new region-group, say $H_j$.
    Then $H_j$ is placed into rack $R_{m_{jN_g}}$, where $m_{jN_g}$ is the $(j, N_g)$-entry of ${\cal M}$, referring to Fig. \ref{fig:rack_layout}(e) as an example.

\end{itemize}

The following two theorems show that
$D^3$ achieves Objective 2 and Objective 3 respectively.

\begin{theorem}
    \label{theorem:recovery-rack-level-1}

    With the data layout of $D^3$,
    the number of cross-rack accessed blocks for recovering a single node failure is minimized for data layouts against a single node failure.

\end{theorem}

\begin{theorem}
    \label{theorem:recovery-rack-level-2}

    $D^3$ achieves load balance among surviving racks in terms of overheads of cross-rack read and write. It also achieves load balance among all nodes in surviving racks in terms of overheads of read, write, and computing respectively, for single-node failure recovery.

\end{theorem}

\vspace{-15pt}
\subsection{Recovery for Locally Repairable Codes}
From \ref{data layout of lrc}, we used the first $N_g$ columns of ${\cal M}$ to guarantee load balance in rack level.
For single node failure recovery, we will choose a new rack to place the reconstructed block for per failed block.
The ids of new racks are chosen based on the last column of  ${\cal M}$.
There are two types of failed blocks:
\begin{itemize}
  \item \textbf{Data blocks/local parity blocks.} For a failed  data block or local parity block, it can be reconstructed by other blocks within the  local
group and  read $\frac{k}{l}$ blocks across racks. To place the reconstructed blocks, we first choose racks based on the last column of the ${\cal M}$, and then choose nodes within the racks in a round-robin way.
The chosen  nodes execute reconstructed tasks and store the obtained reconstructed blocks.
  \item \textbf{Global parity blocks.} When reconstructing a failed global parity block, it needs read all $l+g-1$  parity blocks for reconstruction. The way choosing the targeted racks and nodes
for the reconstructed blocks is the same as the case above.
\end{itemize}
On  one hand, in every region, ${\cal A}$ guarantees balanced recovery traffic in node level.
On the other hand, for load balance of recovery traffic among multiple regions, the last column of ${\cal M}$ gives uniform choices from surviving racks, and
the first $N_g$ columns of ${\cal M}$ guarantee recovery traffic load balance in rack level. Therefore, we get the following theorem:
\begin{theorem}    \label{lrc_recovery}
    When recovering single-node failure, $D^3$ achieves load balance among surviving nodes in terms of read, write,
    and computing for Locally Repairable Codes respectively .
\end{theorem}

\vspace{-15pt}
\subsection{Maintain the original $D^3$ data layout}
$D^3$ achieves minimum cross-rack traffic and optimal load balance in node and rack level  when recovering a single node failure.
So we need do some maintenance for keeping the data layout to guarantee  excellent recovery performance. When one recovery process completes,  the data layout may not keep the group rules, i.e., $t = len \bmod N_g$ groups with $Size_{max}$ blocks and the remaining with $Size_{min}$ blocks (for LRC, $Size_{min}=Size_{max}=1$), and some rack may hold
less blocks than the other racks  because the rack with failed node is not used to   place recovered blocks.
Based on the above two reasons, we get an interim data layout after completing one  recovery.  Before  failures occur again, we need migrate part of data to
the relived node (new online node or new replaced node in the rack with failed node), of course, which   can be delayed   to leisure state of DSS.
In order to minimize the influence on front-end applications in DSSes, we design the migration strategy as following:
\begin{itemize}
  \item \textbf{Batch-by-batch migration.} We need migrate all of the recovered blocks to the relived node to guarantee high recovery performance next time.
  Because storage capacity of a node can get dozens of TB \cite{s4},  we need  migrate massive data,
  which may cause network congestion, packet loss and so on, of course, and the front-end requests also gets performance degradation.
  To minimizing interference on other services, we split the migration operation to multiple batches and migrate the  recovered blocks belonging to  the same type stripe region.
  \item \textbf{Minimum data traffic of one batch.} For each batch of migration, we need to achieve optimal tradeoff between migration speed and   data traffic.
  Fast migration means  massive data traffic in one batch, but heavy cross-rack migration traffic causes scarce source occupation, such as bandwidth, and negative impacts on front-end applications. In each batch,  we  choose all the recovered blocks  in $n-1$  region-groups with recovered blocks of the same type from $n-1$  racks without failed node, which can achieve balanced migration traffic with lowest migration load.
\end{itemize}
For the above migration strategy, we have the following theorem to guarantee the desired properties.

\begin{theorem}
The migration gets optimal tradeoff between migration speed and data traffic,
and data traffic in the process is load balance among $r-1$ racks in DSSes.
\end{theorem}
For an example in Fig. \ref{fig:rack_layout}(e), we execute migration in three batches: $[H_2,H_9,H_{11},H_{18}]$, $[G^{3*}_2,G^{0*}_2,G^{14*}_2,G^{8*}_2]$ and
$[G^{12*}_2,G^{6*}_2,G^{16*}_2,G^{17*}_2]$, and each batch causes $4\times3$ or $4\times6$ blocks cross-rack transferring.
In summary, we achieve load balance in migration to maintain the original $D^3$.

%% file: sections/6-evaluation.tex
\vspace{-5pt}
\section{Performance Evaluation} \label{sec:evaluation}

\subsection{Evaluation Methodology}

We implement $D^3$ in \textit{HDFS Erasure Coding} (HDFS-EC) \cite{b28}
shipped with HDFS in Apache Hadoop 3.1.x.
HDFS-EC is a module integrated in HDFS to support erasure-coded solutions
with built-in erasure coding policies including
the $(2,1)$, $(3,2)$, and $(6,3)$-RS codes.
An HDFS cluster consists of a \textit{NameNode} and a number of \textit{DataNodes}.
The NameNode stores the metadata (e.g., block locations) for managing file operations,
while the DataNodes store data.
In terms of overhead, $D^3$ uses orthogonal arrays to define block placement
and the deterministic distribution of data blocks.
So it only needs to store the orthogonal arrays in the metadata server,
and its memory cost only depends on the number of racks and nodes,
which is around several KB for dozens of nodes.
Besides, other overheads like CPU cost for addressing is negligible
compared to the network transfer cost,
because network bandwidth is usually considered as a scarcer resource in DSSes.

We conduct our evaluation on a cluster of 28 machines, each of which is configured with a quad-core 3.40 GHz Intel Core i5-7500, 8 GB memory, and a Seagate ST1000DM010-2EP102 7200 RPM 1 TB SATA hard disk,
and runs Ubuntu 16.04.
One machine runs as the NameNode and each of the other machines runs as a DataNode.
All DataNodes are divided into some racks, where all machines within a rack are connected via a UTT SG108V 8-Port 1000 Mb/s Ethernet switch and all racks are connected via a D-LINK DES-1024D 24-Port 100 Mb/s Ethernet switch to simulate practical systems.
The available cross-rack bandwidth per node is only 1/20 to 1/5 of the inner-rack bandwidth in practical systems \cite{b14, e2}, and more specifically, the available inner-rack bandwidth per node is 1000 Mb/s as described in \cite{e2}. Therefore, the settings of the inner-rack and cross-rack bandwidths in our experiments are reasonable.

We take the commonly used 
\textit{random data distribution} (RDD)
in practical systems as baseline.
In our experiments, we make RDD randomly distribute blocks of each stripe among all nodes,
while ensuring single-rack fault tolerance.
The recovery solution of RDD is that it randomly chooses $k$ surviving blocks of a stripe
and sends them to a randomly selected node excluding the nodes containing the blocks of the same stripe for recovery.
We write 1000 stripes of blocks with $D^3$ and RDD respectively,
and conduct our evaluation with different settings.
To evaluate the recovery performance,
we randomly pick one DataNode and erase all of its blocks,
and then measure the \textit{recovery throughput},
defined as the total volume of failed blocks being repaired over the total recovery time.
The presented results are averaged over five runs.
%
%

\vspace{-5pt}
\subsection{Experimental Results}


By default, the DSS consists of eight racks with three DataNodes each, and the block size is set as 16 MB using $(2,1)$-RS code. We also conduct experiments with different erasure codes and rack configurations.

\vspace{-5pt}
\subsubsection{Recovery performance}
\noindent \textbf{Experiment 1 (Repair Load Balance).} \hspace{0ex}
Because the cross-rack data transfer is crucial to the system performance, we first evaluate the distributions of cross-rack repair traffic loads of RDD and $D^3$.
In our experiments, each port of the switch connecting the racks is full-duplex,
with 100 Mb/s upstream and 100 Mb/s downstream available simultaneously.
For the port connecting to the $i$-th surviving rack ($0 \leq i \leq 6$),
let $L_i, L'_i$ be the upstream load from and the downstream load to this port respectively.
Let $L_{max} = \max_{0 \leq i \leq 6} \{L_i, L'_i\}$ and
$L_{avg} = \sum_{0 \leq i \leq 6}(L_i + L'_i)/14$.
Then we measure the \textit{load imbalance} by $\lambda = (L_{max} - L_{avg})/L_{avg}$.

\begin{figure}[!htbp]
\setlength{\abovecaptionskip}{5pt}
\setlength{\belowcaptionskip}{-5pt}
    \centering
    \begin{minipage}[t]{0.24\textwidth}
        \centering
        \includegraphics[width=1\textwidth]{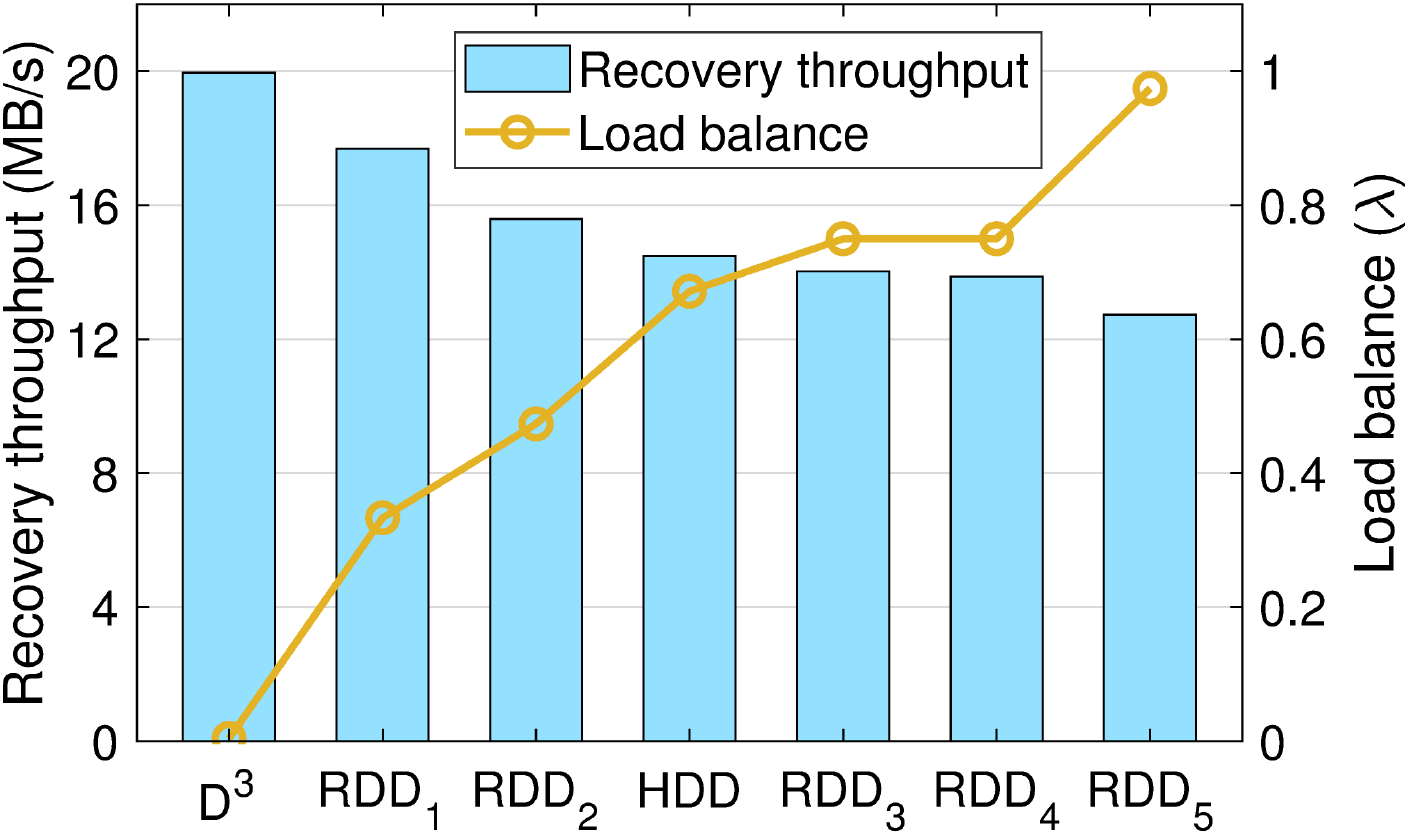}
        \caption{  Recovery under RS.}
        \label{fig:exp1}
    \end{minipage}
    \begin{minipage}[t]{0.24\textwidth}
        \centering
        \includegraphics[width=1\textwidth]{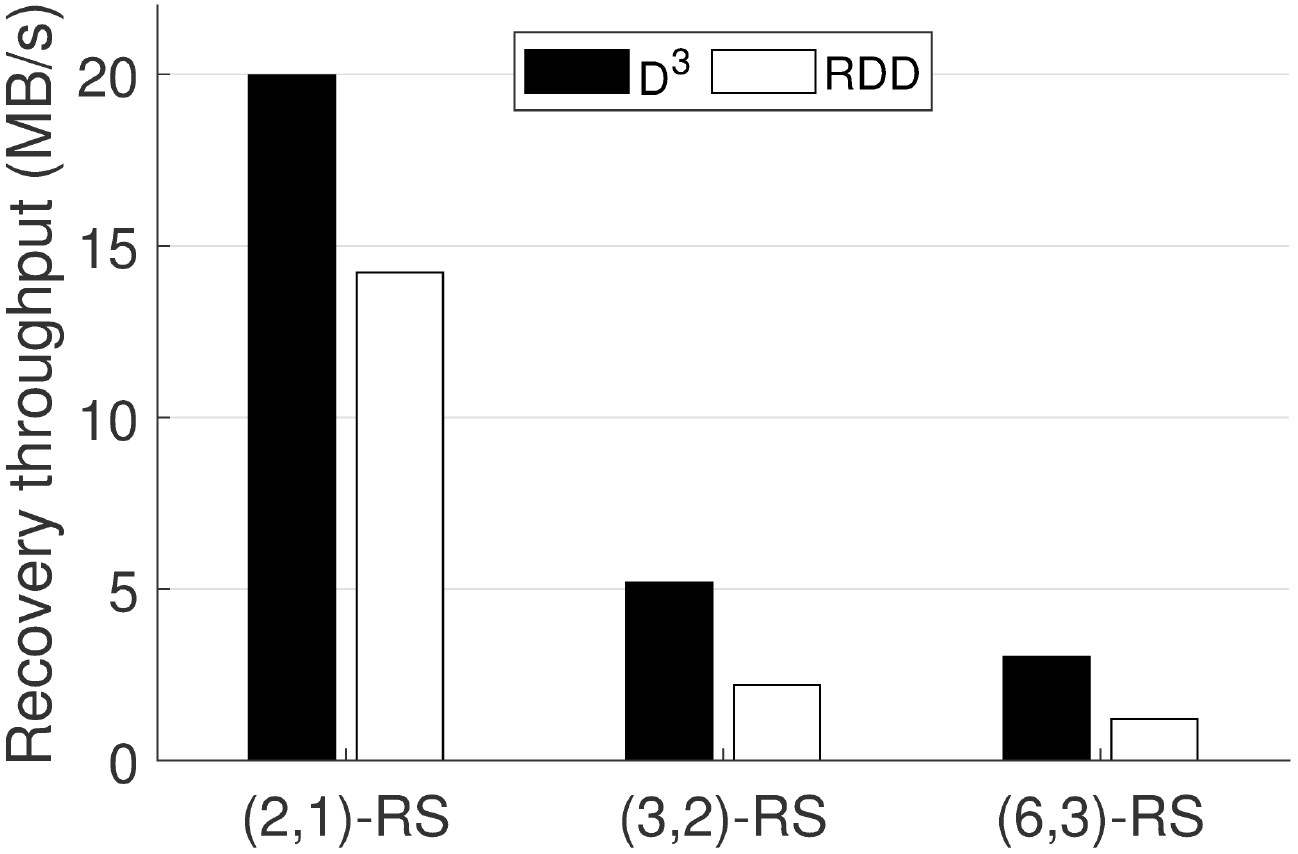}
        \caption{RS codes configuration.}
        \label{fig:exp2}
    \end{minipage}
\end{figure}

We run RDD with five groups,
and each group has a different random data distribution with a randomly chosen failed node.
In each group, we run the experiments five times with the same data distribution and the same failed node, and average the recovery throughput and the corresponding load imbalance metric $\lambda$, as shown in Fig. \ref{fig:exp1},
where $RDD_i$ is the $i$-th group and the results are sorted by $\lambda$.
From Fig. \ref{fig:exp1}, we find the repair load with $D^3$ is balanced due to our deterministic data layout, but it is constantly imbalance with RDD. In the five groups of experiments, $\lambda$ varies from 0.33 to 0.97, i.e., $L_{max}$ is at least 1.33 times of $L_{avg}$, which shows severe imbalance of repair load.
RDD achieves uniform distribution of blocks as long as the number of stripes is large enough. However, due to the limited resources in a node, such as memory  and CPU, etc, the reconstruction is executed batch by batch. As a result, the distribution of blocks with RDD is severely skewed in a batch which has  limited number of stripes.
We can also find from Fig. \ref{fig:exp1} that the repair load imbalance significantly slows down the recovery process and the recovery throughput decreases when $\lambda$ increases. $D^3$ increases the recovery throughput by 35.92\% on average,
compared with RDD.

We also evaluated the \textit{hash-based data distribution} (HDD) for comparison.
HDD uses a pseudo-random hash function to
map an input value (typically a block or data object identifier)
to a list of storage devices,
and it generates a pseudo-random (but deterministic) data distribution \cite{r11, b21, r5}.
In the implementation of HDD,
we use the Jenkins hash function \cite{e1},
which is adopted by data placement algorithms such as CRUSH \cite{b21},
to map blocks to nodes,
while ensuring single-rack fault tolerance.
HDD will reselect nodes using a modified input
(by following reselection behavior of CRUSH \cite{b21})
for three different reasons:
(1) if a block is mapped to a node already used in the current stripe,
(2) if a block is mapped to a node which cannot ensure single-rack fault tolerance, or (3) if a node is failed.
The recovery throughput and the corresponding load imbalance metric $\lambda$ of HDD
are also shown in Fig. \ref{fig:exp1}, where marked as ``HDD''.
$D^3$ increases the recovery throughput by 37.83\%, compared with HDD.

\noindent \textbf{Experiment 2 (Erasure-coded Configuration).} \hspace{0ex}
We use three HDFS build-in $(2,1)$, $(3,2)$, and $(6,3)$-RS codes to evaluate the recovery throughput of $D^3$ with different erasure code settings, shown in Fig. \ref{fig:exp2}.
We find that the recovery throughput decreases as the stripe size increases from
$(2,1)$, $(3,2)$ to $(6,3)$-RS.
This is because more blocks participate in recovering one failed block with larger stripe size,
which induces heavier load and thus reduces the recovery throughput.
We also find that $D^3$ outperforms RDD more significantly as the stripe size increases from $(2,1)$, $(3,2)$ to $(6,3)$-RS code. The recovery throughput of $D^3$ with $(2,1)$, $(3,2)$, and $(6,3)$-RS codes are 1.40, 2.36, and 2.49 times of which of RDD respectively.
This is because with more blocks in a rack as $(3,2)$ and $(6,3)$-RS codes,
$D^3$ aggregates more blocks into an aggregated one and saves more cross-rack traffic.

\begin{figure}[!htbp]
\setlength{\abovecaptionskip}{5pt}
\setlength{\belowcaptionskip}{-5pt}
    \centering
        \begin{minipage}[t]{0.24\textwidth}
        \centering
        \includegraphics[width=1\textwidth]{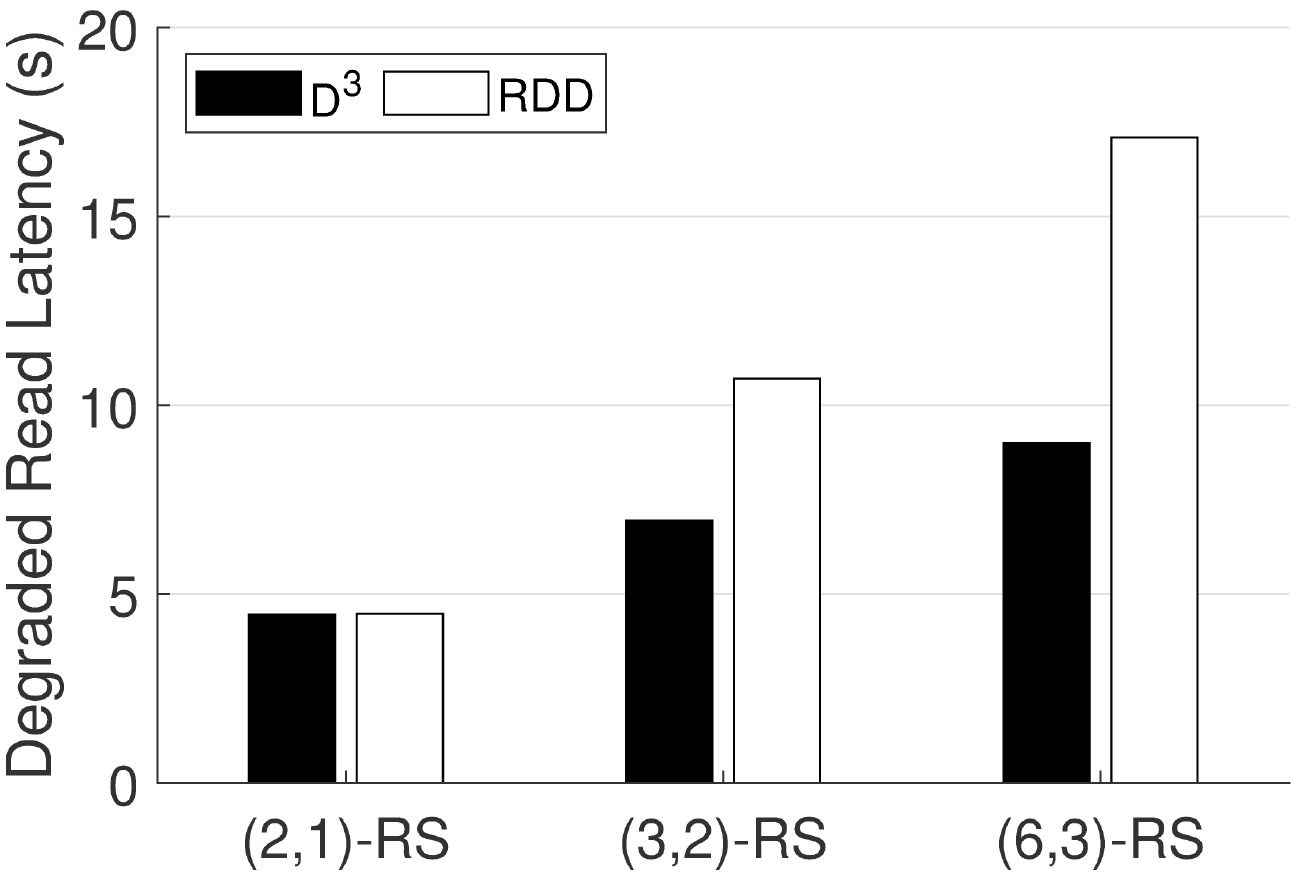}
        \caption{ Degraded read latency.}
        \label{fig:exp7a}
    \end{minipage}
    \begin{minipage}[t]{0.24\textwidth}
        \centering
        \includegraphics[width=1\textwidth]{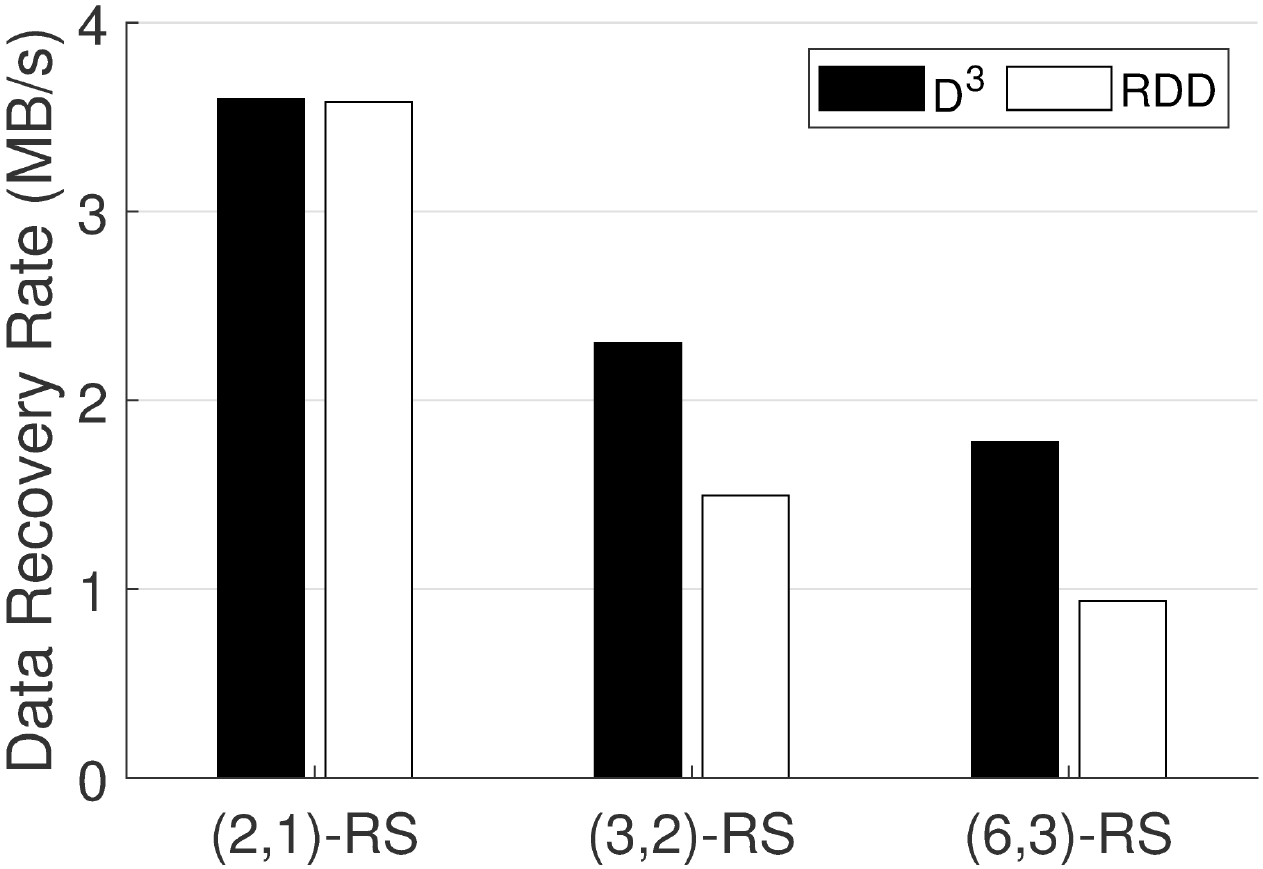}
        \caption{Data recovery rate.}
        \label{fig:exp7b}
    \end{minipage}
\end{figure}

\noindent \textbf{Experiment 3 (Degraded Read).} \hspace{0.5ex}
A client (i.e., a HDFS node), which tries to read a lost data block that
has not been repaired, will trigger
a \textit{degraded read} to repair the unavailable block.
To evaluate the performance of degraded read,
we randomly erase one data block and randomly choose a client to read the block,
then measure the \textit{degraded read latency},
defined as the time from issuing a read request
until the failed block is repaired at the client.
Fig. \ref{fig:exp7a} shows the degraded read latency for $D^3$ and RDD
with $(2,1)$, $(3,2)$, and $(6,3)$-RS codes.
We find that the degraded read latencies of $D^3$ and RDD are
almost identical for $(2,1)$-RS code.
This is because each block of a single $(2,1)$-RS stripe is placed in a distinct rack
to achieve rack-level fault tolerance for both $D^3$ and RDD.
With $(3,2)$ and $(6,3)$-RS codes,
the degraded read latencies of $D^3$
are reduced by 35.16\% and 47.34\% respectively,
compared with RDD.
This is because $D^3$ aggregates blocks within racks
and minimizes cross-rack accessed blocks within a single stripe.
So $D^3$ induces less cross-rack traffic for recovery
and has a shorter degraded read latency.

We also plot the \textit{data recovery rate},
defined as the size of the failed block divided by the degraded read time,
as shown in Fig. \ref{fig:exp7b}.
We find that,
for each of $(2,1)$, $(3,2)$, and $(6,3)$-RS codes,
the data recovery rate in Fig. \ref{fig:exp7b}
is smaller than
the recovery throughput in Fig. \ref{fig:exp2},
especially much smaller for $(2,1)$-RS code.
This is because degraded read only repairs a single block,
while node recovery needs to repair many blocks, which induces more racks and nodes participating in the recovery with a larger degree of parallelism.

\vspace{-5pt}
\subsubsection{Sensitivity to Internal Parameters}

\begin{figure}[!htbp]
\setlength{\abovecaptionskip}{5pt}
\setlength{\belowcaptionskip}{-5pt}
    \centering
        \begin{minipage}[t]{0.24\textwidth}
        \centering
        \includegraphics[width=1\textwidth]{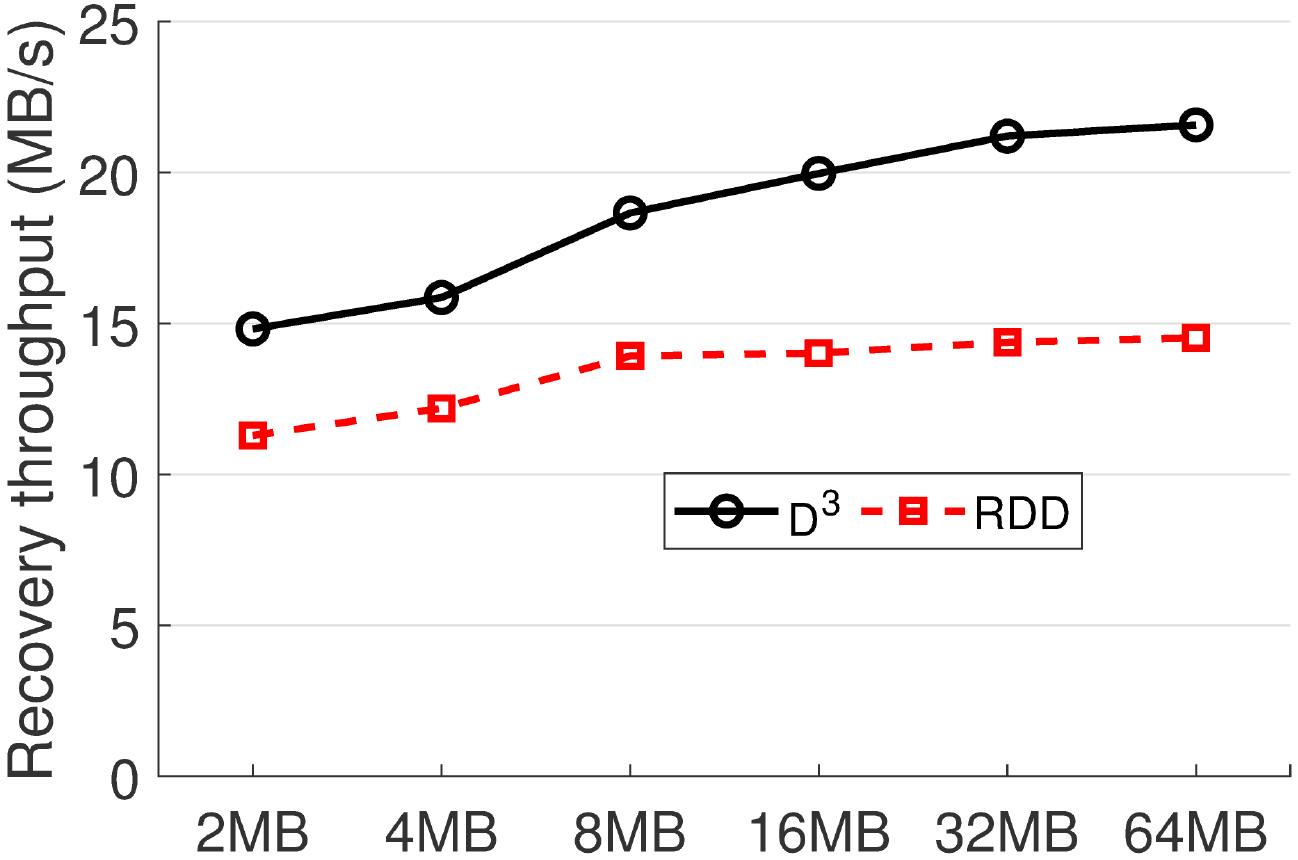}
        \caption{Block size.}
        \label{fig:exp3}
    \end{minipage}
    \begin{minipage}[t]{0.24\textwidth}
        \centering
        \includegraphics[width=1\textwidth]{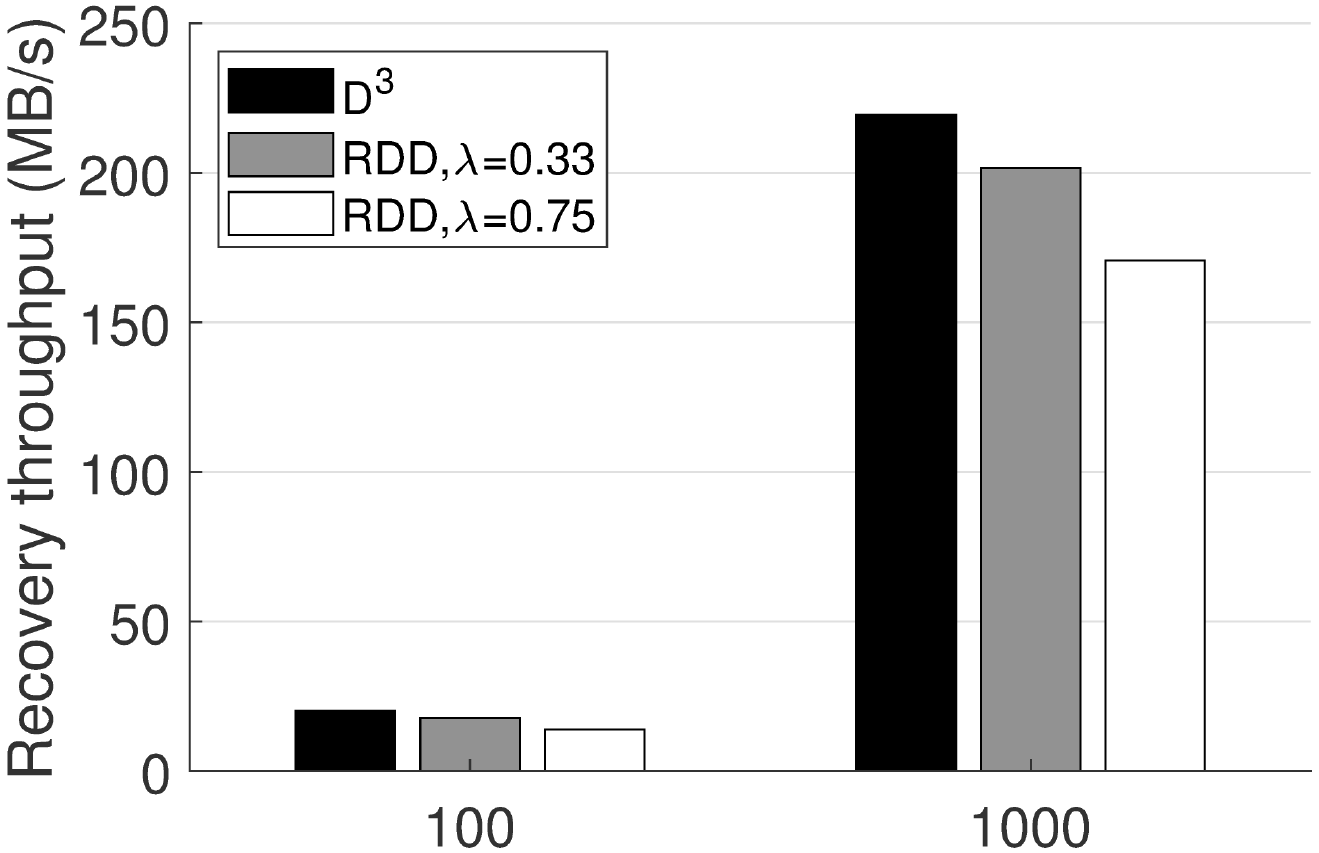}
        \caption{Cross-rack bandwidth.}
        \label{fig:exp4}
    \end{minipage}
\end{figure}

 \noindent \textbf{Experiment 4 (Block Size).} \hspace{0ex}
We now study the performance of $D^3$ with different block sizes.
We fix the distribution of RDD with $\lambda = 0.75$ and
change the block size.
Fig. \ref{fig:exp3} shows the recovery throughput
for the block sizes from 2 to 64 MB.
We find that with both $D^3$ and RDD, the recovery throughput almost increases along with the increase of block size,
which reaches the maximum when the block size is 32 MB.
But compared to RDD, $D^3$ increases the recovery throughput almost with a consistent ratio, about 39.57\%, with different block sizes.

\noindent \textbf{Experiment 5 (Cross-Rack Bandwidth.)} \hspace{0ex}
We now present the performance of $D^3$ with different cross-rack bandwidths.
We fix the distribution of RDD with $\lambda = 0.33$ and 0.75, and
change the cross-rack bandwidth
by connecting the racks via
a D-LINK DES-1024D 24-Port 100 Mb/s Ethernet switch or
a TP-LINK TL-SG1016DT 16-Port 1000 Mb/s Ethernet switch.
Fig. \ref{fig:exp4} shows the recovery throughput.
On average, $D^3$ increases 27.82\% and 18.10\% of the recovery throughput
compared to RDD with cross-rack bandwidth of 100 Mb/s and 1000 Mb/s, respectively.
Furthermore, we find that the recovery throughput increases significantly
when the cross-rack bandwidth increases from 100 Mb/s to 1000 Mb/s,
which implies that the cross-rack traffic is crucial to the recovery performance.

\begin{figure}[!htbp]
\setlength{\abovecaptionskip}{5pt}
\setlength{\belowcaptionskip}{-5pt}
    \centering
        \begin{minipage}[t]{0.24\textwidth}
        \centering
        \includegraphics[width=1\textwidth]{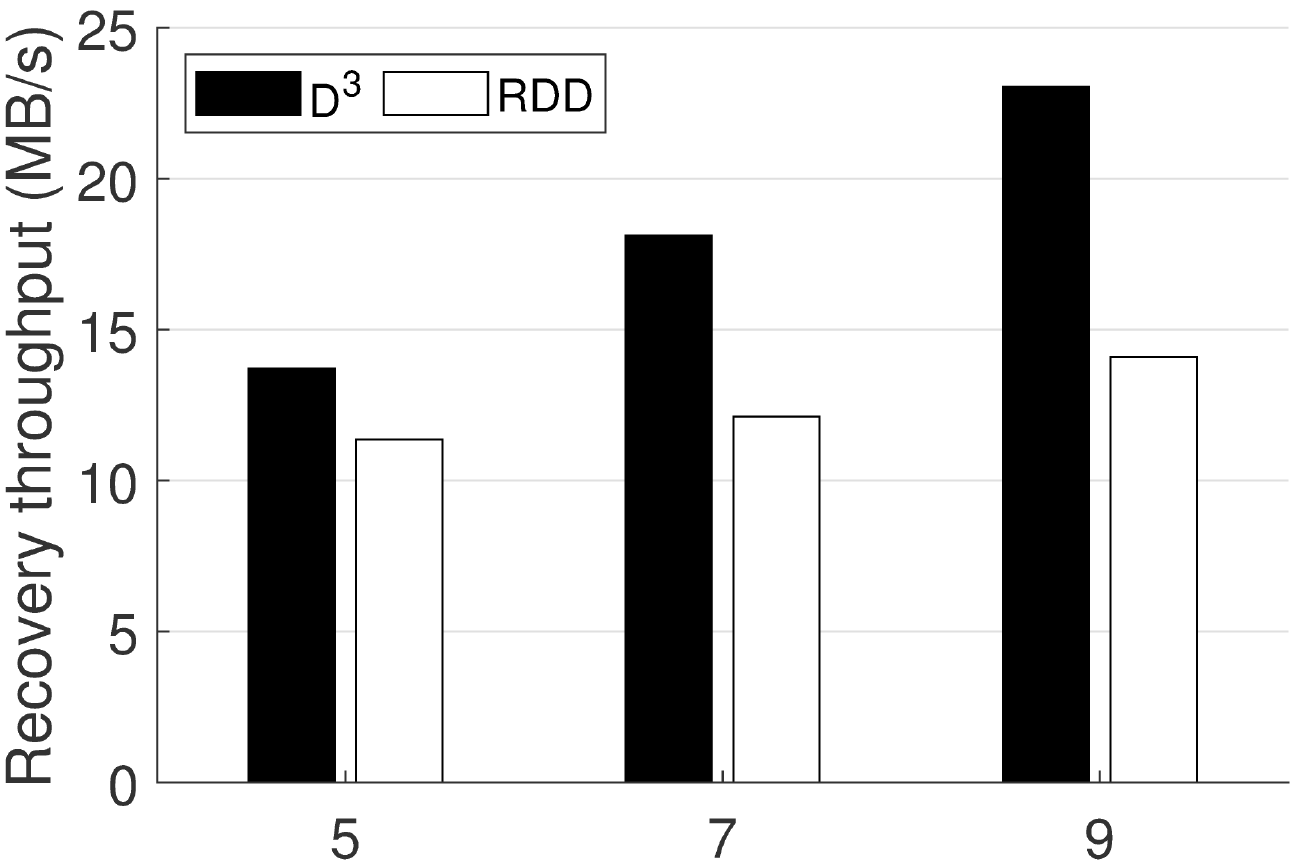}
        \caption{Number of racks.}
        \label{fig:exp5}
    \end{minipage}
    \begin{minipage}[t]{0.24\textwidth}
        \centering
        \includegraphics[width=1\textwidth]{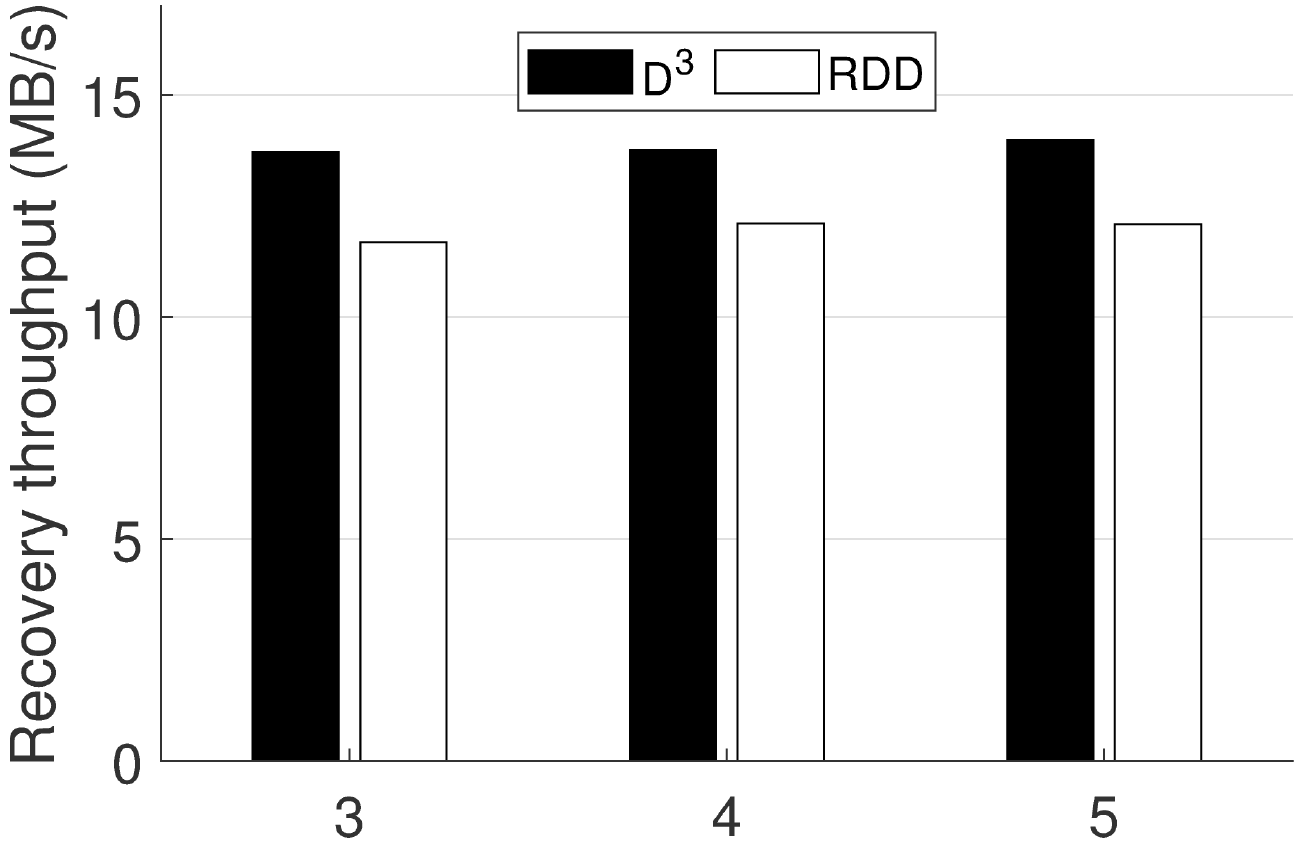}
        \caption{Number of nodes per rack.}
        \label{fig:exp6}
    \end{minipage}
\end{figure}

\noindent \textbf{Experiment 6 (Number of Racks).} \hspace{0ex}
We fix the number of nodes in each rack as three to evaluate the performance of $D^3$
with five, seven, and nine racks, as shown in Fig. \ref{fig:exp5}.
We find that the recovery throughput increases
with more racks.
This is because the cross-rack bandwidth is crucial to recovery throughput,
and the total cross-rack bandwidth increases with more racks, which benefits the failure recovery and increases the recovery throughput.
We also find that $D^3$ increases the recovery throughput more evidently with more racks,
1.21, 1.49, and 1.64 times of which of RDD with five, seven, and nine racks respectively.
This is because with more racks, RDD suffers more severe imbalance of repair load.

\noindent \textbf{Experiment 7 (Number of Nodes per Rack).} \hspace{0.5ex}
We fix the number of racks as five, and set three, four, and five nodes per rack to evaluate the performance of $D^3$, as shown in Fig. \ref{fig:exp6}.
When the number of nodes per rack varies from three to five, the recovery throughput with $D^3$ and RDD does not significantly vary.
The reason is that
the cross-rack bandwidth,
which is crucial to the recovery throughput,
will not change when the number of nodes per rack varies.

\vspace{-5pt}
\subsubsection{Recovery performance of LRCs}
In this experiment,
we study the performance of $D^3$ for Locally Repairable Code (LRC) \cite{b18}.
The hadoop 3.1.x currently does not support LRC policies, but we implement and mimic a representative LRC policy in HDFS,
which is $(4,2,1)$-LRC in the previous example, and we use it to compare the recovery throughput of $D^3$ and RDD.
$D^3$ uses OA$(3,3)$ and OA$(8, 4)$ to address node-level locations and place regions in rack level, respectively.

\begin{figure*}[!t]
\setlength{\abovecaptionskip}{5pt}
\setlength{\belowcaptionskip}{-10pt}
    \centering
        \begin{minipage}[t]{0.24\textwidth}
        \includegraphics[width=1\textwidth]{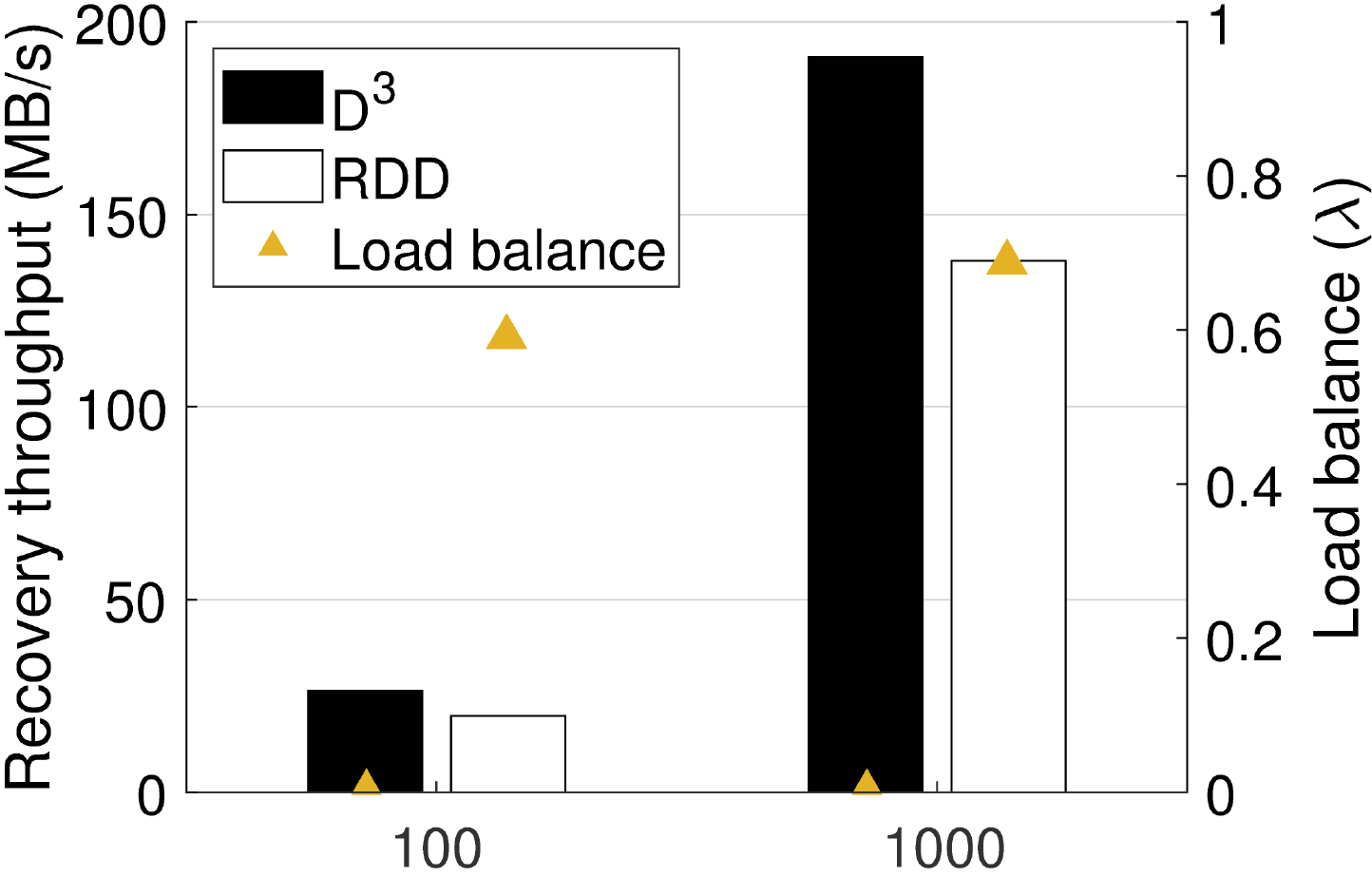}
        \caption{ Recovery under LRC.}
        \label{fig:exp8}
    \end{minipage}
    \begin{minipage}[t]{0.24\textwidth}
        \includegraphics[width=1\textwidth]{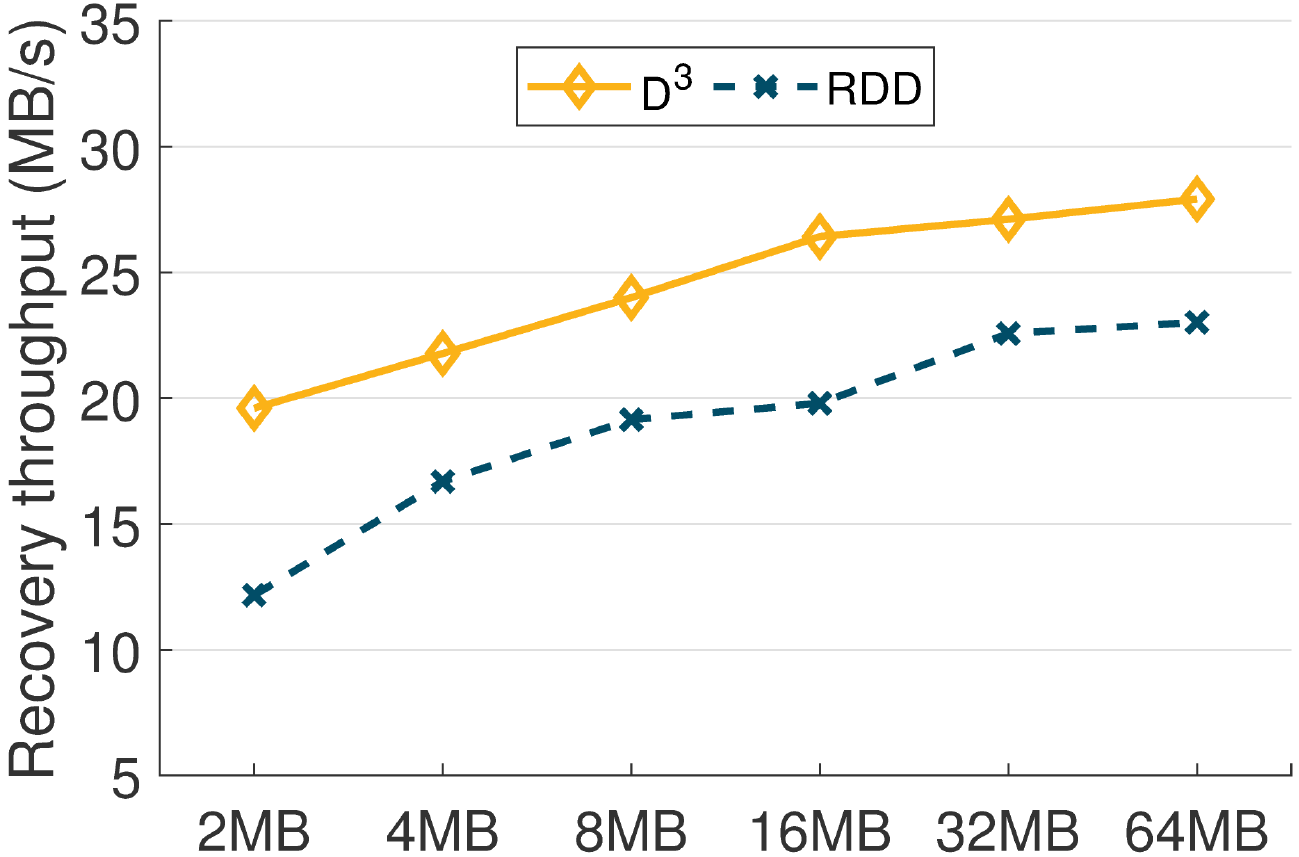}
        \caption{Block size under LRC.}
        \label{fig:exp11}
    \end{minipage}
    \begin{minipage}[t]{0.24\textwidth}
        \centering
        \includegraphics[width=1\textwidth]{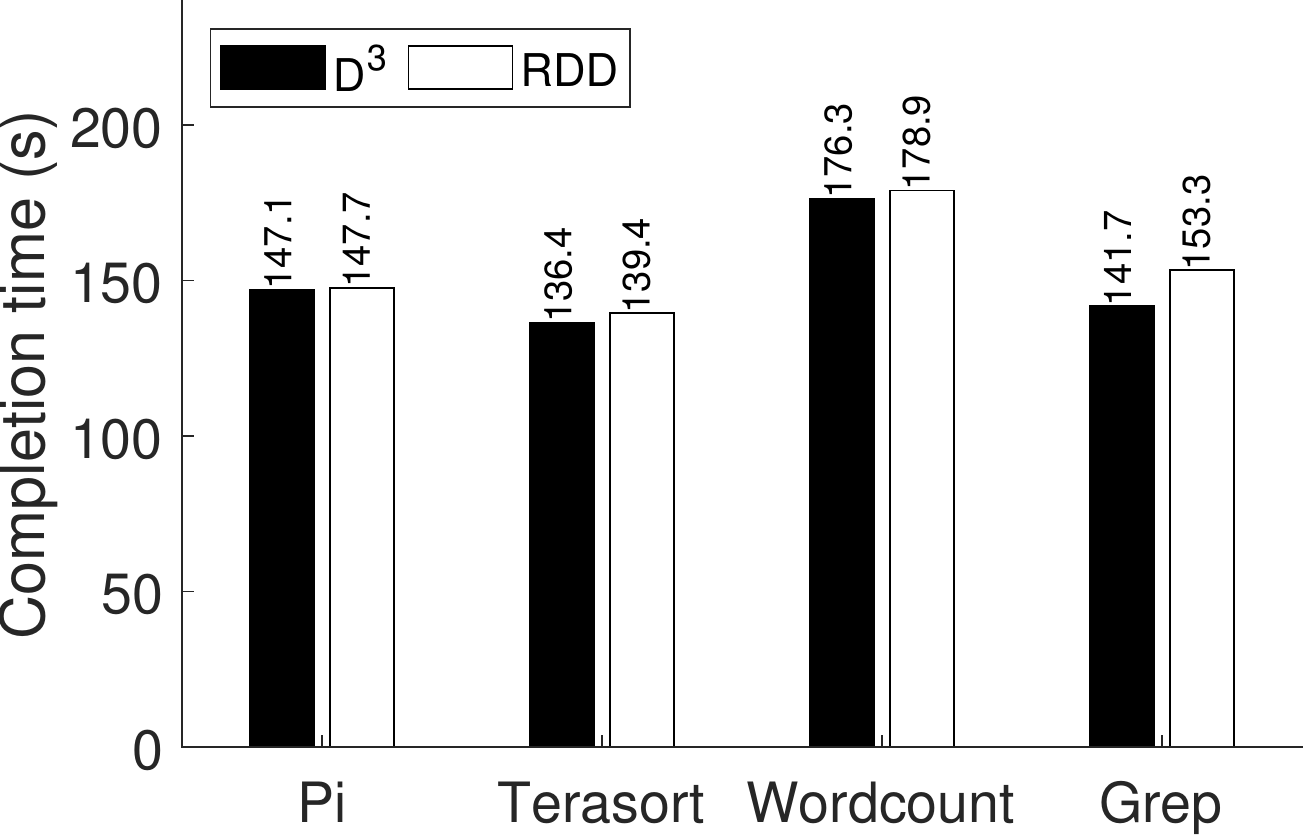}
        \caption{Normal state.}
        \label{fig:exp9}
    \end{minipage}
    \begin{minipage}[t]{0.24\textwidth}
        \centering
        \includegraphics[width=1\textwidth]{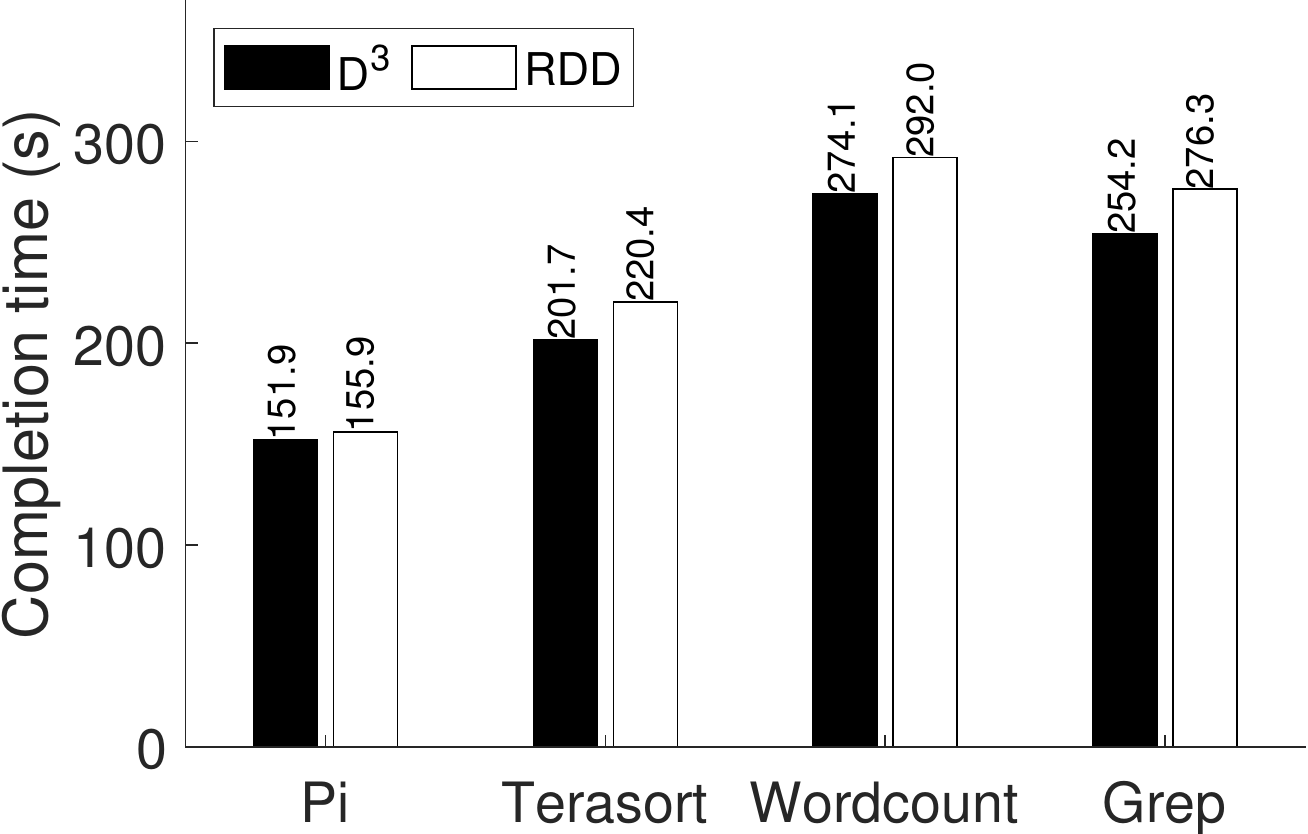}
        \caption{Recovery state.}
        \label{fig:exp10}
    \end{minipage}
\end{figure*}

\noindent \textbf{Experiment 8 (Recovery performance under LRC).}
Fig. \ref{fig:exp8} shows the recovery throughput of $D^3$ and RDD
over $(4,2,1)$-LRC with different cross-rack bandwidth configuration.
When the cross-rack bandwidth is 100 Mb/s and 1000 Mb/s and
the load balance $\lambda$s of RDD are $0.5909$ and $0.6879$ respectively.
$D^3$ increases 40.23\% and 38.35\% of the recovery throughput
compared with RDD, respectively.
The most primary reason is that $D^3$ achieves  uniform distribution of recovery
traffic, while random network traffic distribution of RDD in recovery causes
the network-bandwidth proportions of rack-level switch ports are unbalanced.
From another point of view, $D^3$ also balances computing tasks to different nodes, so
it can also uniformly allocate CPU source and memory usage among surviving nodes.

\noindent \textbf{Experiment 9 (Impact of block size under LRC).}
Block sizes are different for diverse users' demand and document types.
In Fig. \ref{fig:exp11}, we first take fixed $\lambda = 0.5909$ for RDD,
and then we study the impact of block size in recovery under $D^3$ and RDD. With the increase  of block size, the recovery throughput of $D^3$ and RDD both increases.
Compared with RDD, $D^3$ increases recovery throughput over LRC by $20.13\% \sim 61.10\%$, and 31.98\% on average.
The reason is that reconstructed tasks for $D^3$ are uniformly distributed in node level,
causing cross-rack recovery traffic uniformly distributed among racks.

\vspace{-5pt}
\subsubsection{Benchmark tests performance}
A DSS also usually needs support many front-end users' responses, whose performance needs to be prioritized.
In the experiment, we evaluate four hadoop benchmarks, namely $Pi$, $Terasort$, $Wordcount$ and $Grep$, which will generate and store temporary data
in HDFS depending on data layout in the tasks execution phase.
The failure recovery of DSSes mainly competes cross-rack network bandwidth and CPU source with the four benchmarks.
We still take the representative $(2,1)$-RS code as storage policy.
The following describes characteristics of the four workloads, and the detailed configuration is shown in Table \ref{tab:application}.
\begin{itemize}
  \item \textbf{Pi.} The map/reduce program uses a BBP-type method  \cite{r21} to compute exact binary digits of $\pi$,
  the approximation algorithm needs lots of computing source,  and the more tasks execution times, the more precise numerical value of $\pi$ \cite{r22}.
  \item \textbf{Terasort}. It is a standard map/reduce sort, except for a custom partitioner that uses a sorted list of N-1 sampled keys to define the key range for each reduce.
  It first samples the input and computes the input
distribution, and then writes the list of keys into HDFS. TeraGen command
generates the large random input data for Terasort \cite{r19}.
  \item \textbf{Wordcount}.
 It reads the text input files, breaks each line into words
  and counts them. The output is a locally sorted list of words and the
  counts of how often they occurred \cite{r19,r20}.
  \item \textbf{Grep}. It is an example of Hadoop Map/Reduce application. It extracts matched strings provided by users from text
files and sorts matched strings by their frequency \cite{r19}.
\end{itemize}

\begin{table*}[!htbp]
\setlength{\belowcaptionskip}{10pt}
\centering
\caption{Hadoop benchmarks.}
\begin{tabular}{cccc}
\hline
Workloads & Application domain & Data source & Configuration\\
\hline
Pi & CPU intensive & Autogeneration & 100 maps with 100m samples\\
\hline
Terasort & CPU/Network intensive & Table & Generating 5m records by TeraGen\\
\hline
Wordcount & Network intensive & Text & Randomly generating 100m words\\
\hline
Grep & Network intensive & Text & Randomly generating 100m words\\
\hline
\end{tabular}
\label {tab:application}
\end{table*}

\vspace{-5pt}
\noindent \textbf{Experiment 10 (Benchmark tests in normal).}
From Fig. \ref{fig:exp9}, we find $D^3$ shows better support to front-end users' responses when running the four workloads.
The reason is that $D^3$ achieves a uniform data distribution for the intermediate data when running  MapReduce tasks,
which benefits distribution of network traffic when accessing temporarily stored data across nodes in HDFS.
More specifically, for Pi, it mainly occupies CPU source of DataNodes and involves little network traffic, so $D^3$ doesn't show
significant advantage when compared with RDD.
Terasort and Wordcount  need more cross-node network traffic, so more advantage is shown better because $D^3$ stores
more uniform data.
For Grep, which frequently needs extract and sort matched strings,  it causes heavy cross-node  traffic, so the system performance based on $D^3$ is improved  up to  $7.57\%$.

\noindent
\textbf{Experiment 11 (Benchmark tests in recovery).}
We first write $3000$ stripes and then randomly erase one node to mimic single node failure.
In the next moment, we run the mapreduce tasks with the same configuration in Table \ref{tab:application}
from Namenode, and
task execution process of the four benchmarks is in recovery state of DSSes.
In Fig. \ref{fig:exp10}, the recovery of $D^3$ only causes $3.26\%$ performance reduction when compared
with normal-state performance for Pi. The reason is that $D^3$ uniformly distributes reconstruction tasks to computing nodes with uniform CPU and memory occupation.
For network-intensive Terasort, Wordcount and Grep,
because of balanced network bandwidth among racks and nodes,
$D^3$ reduces completion time by $8.48\%$, $6.13\%$ and $8.00\%$ respectively when compared with RDD.

%% file: sections/7-related-work.tex
\section{related work} \label{sec:related_work}

\noindent \textbf{Data Layout Optimization.} \hspace{0.5ex}
Data layout refers to how to place data/parity blocks or replicas across disks or servers, which plays a critical role in both performance and reliability
of RAIDs and DSSes.
There are some approaches to improve RAID performance
by means of data layouts \cite{r1, r2, b32, r3}.
Parity declustering was first proposed
by Muntz and Lui \cite{r1} as a data layout technique
to speed up the failure recovery process and provide
highly-available arrays.
It requires less
additional load on surviving disks for data reconstruction
and the rebuild read load is balanced among all surviving disks.
Parity declustering was further realized by Holland and Gibson \cite{r2} based
on balanced incomplete block designs in practical systems.
Although parity declustering spreads rebuild reads to surviving disks,
rebuild is still capped by the write speed of the replacement disk.
RAID+ \cite{b32} spreads both rebuild read and write to the surviving disks, and as a result, it speeds up data recovery for RAIDs.
However, the data layouts of RAID+ for RAIDs are not efficient for failure recovery in DSSes. Unlike RAIDs, where the organization of disks is flat,
DSSes typically connect storage nodes with a tree-structured topology.
Therefore DSSes often
exhibits the property of heterogeneous bandwidth,
where the available inner-rack bandwidth is considered to be sufficient,
while cross-rack bandwidth is often over-subscribed.
Furthermore, in DSSes, each server has both storage resources and computing power, so every server can participate in
 encoding/decoding, while for RAIDs, computing can only be performed  on the server that connecting the disks.

For data layouts in DSSes,
EAR \cite{b17} is a placement algorithm
for efficient transfer from replication to erasure coding.
There are also some pseudo-random hash-based data placement approaches for scalable storage systems.
Brinkmann et al. \cite{r5} use pseudo-random hash-functions for evenly
distributing and efficiently locating data in dynamically changing SANs.
Chord \cite{r6} uses a variant of consistent hashing \cite{r7}
to assign keys to storage nodes
for efficient adaption to dynamic change of nodes in systems.
SCADDAR \cite{r8} uses a pseudo-random placement to distribute continuous media
blocks across all disks, and it minimizes block movement in case of disk scaling.
RUSH \cite{r11} and CRUSH \cite{b21} utilize
pseudo-random hash-based functions to
map replicated objects to a scalable collection of storage devices.
They are probabilistically optimal in distributing data evenly and
minimizing data migration when new storage devices are added to systems.
These hash-based data placement strategies are
designed for improving scalability in dynamic distributed environments.
They all use pseudo-random hash functions to generate pseudo-random data distribution,
while $D^3$ presents a deterministic data distribution to speed up data reconstruction.

\vspace{-5pt}
\vspace{1ex} \noindent \textbf{Erasure Codes for Fast Recovery.} \hspace{0.5ex}
RS codes \cite{b9} are the most popular erasure codes
that are widely deployed in real storage systems \cite{b1, b2, b10}.
Rotated RS codes \cite{r13} convert the one-row RS codes into multi-row codes
and optimize the coding chains, which induces fewer data reads for faster degraded read.
Regenerating codes \cite{r14} minimize the repair traffic by allowing
other surviving nodes to send computed data for data
reconstruction, and achieve an optimal tradeoff between
storage redundancy and repair
traffic. In particular, minimum-storage regenerating (MSR)
codes \cite{r14} are MDS, and they minimize the repair traffic subject to the minimum storage redundancy.
Rashmi et al. \cite{r15} propose a new MSR code construction
that also minimizes the recovery I/Os.
A class of erasure codes called LRCs \cite{b18, b7} are proposed to
reduce the repair traffic, with
at least 25\% to 50\% additional parities,
and hence they are non-MDS.
HACFS \cite{r18} dynamically switches between two different erasure codes
according to workload changes
so as to balance storage overhead and recovery performance.
Our $D^3$ is a data placement scheme which deterministically distributes blocks into DSSes based on erasure codes to improve repair performance.

%% file: sections/8-conclusion.tex
\vspace{-10pt}
\section{conclusion} \label{sec:conclusion}

This paper proposes $D^3$,
a \textit{deterministic data distribution} scheme which
significantly speeds up failure recovery.
It defines the data distribution with orthogonal arrays, which reaches even distribution of data/parity blocks among nodes, balanced repair traffic among nodes and racks, and the minimal cross-rack repair traffic against a single node failure.
It provides good support to MDS codes and LRCs.
Experimental results show that $D^3$ significantly speeds up the failure recovery process and
efficiently supports front-end users' applications,
when compared with random data distribution.

%% file: sections/9-appendix.tex
\section*{appendix} \label{sec:appendix}

\subsection*{Proofs for Results in Section 4}

\noindent \textbf{Proof of Lemma 1.}
This is because $Size_{max} = \lceil len/N_g \rceil = \lceil len/ (\lceil len / m \rceil) \rceil \leq \lceil len/ ( len / m ) \rceil = m$.
\hfill \qed 

\noindent \textbf{Proof of Lemma 2.}
If there is no group containing no more than $m-1$ blocks,
then each group contains exactly $m$ blocks by Lemma \ref{lemma:stripe-level-1},
which leads to $b=0$, contradicting to $b > 0$.

If there is only one group containing no more than $m-1$ blocks,
then any other group contains $m$ blocks.
So we have $Size_{max} = \lceil len/N_g \rceil = m$, and further $Size_{min} = \lfloor len/N_g \rfloor = m-1$.
We have $b = m-1$, which contradicts to $b < m-1$.

Thus, if $0 < b < m-1$, there are at least two groups
with each group containing no more than $m-1$ blocks.
\hfill \qed 

\noindent \textbf{Proof of Lemma 3.}
Denote the $k$-th block in stripe $S_i$ as $B_{i,k}$ for $0 \leq k \leq len, 0 \leq i \leq n^2-1$. Suppose $B_{i,k}$ is the $k'$-th block in the $j$-th group of $S_i$, where $k'$ is the same one for all $0 \leq i \leq n^2-1$.
From Definition \ref{def:oa},
we know that, for a given $j$, each entry $y$ ($0 \leq y \leq n-1$) appears exactly $n$ times
in all $a_{ij}$'s  for $0 \leq i \leq n^2-1$, which form the $j$-th column of ${\cal A}_n$.
So each entry $y$ ($0 \leq y \leq n-1$) appears exactly $n$ times
in the $((a_{ij} + k') \bmod n)$'s ($0 \leq i \leq n^2-1$), which means $n$ blocks from $B_{0,k}, B_{1,k},..., B_{n^2-1,k}$ are placed to $N_{j,y}$ in $R_j$. So data/parity blocks are evenly distributed among the $n$ nodes in each rack.
\hfill \qed 

\noindent \textbf{Proof of Theorem 2.}
We know that region-group $G^i_j$ ($0 \leq i \leq r(r-1)-1, 0 \leq j \leq N_g-1$)
is placed in rack $R_{m_{ij}}$.
According to the definition of ${\cal M}$ and Property \ref{def:pro1} of OA,
given $j$ for $0 \leq j \leq N_g-1$, each entry $y$ ($0 \leq y \leq r-1$) appears exactly $r-1$ times
in the $m_{ij}$'s for $0 \leq i \leq r(r-1)-1$, which form the $j$-th column of ${\cal M}$. So $r-1$ region-groups from $G^0_j, G^1_j, \cdots, G^{r(r-1)}_j$ are placed in rack $R_y$, and further for each $j$ ($0 \leq j \leq N_g-1$),
the $r(r-1)$ region-groups $G^i_j$'s ($0 \leq i \leq r(r-1)-1$) are evenly distributed among all racks.
By Lemma \ref{lemma:node-level},
the data/parity blocks in each $G^i_j$
are evenly distributed among nodes in a rack.
Thus, each node contains the same number of data/parity blocks.
\hfill \qed 

\noindent \textbf{Proof of Theorem 3.}
Since $D^3$ places the blocks within the same stripe to different nodes,
it tolerates $m$ node failures.
Furthermore, by Lemma \ref{lemma:stripe-level-1}, at most $m$ blocks of a stripe are placed into the same rack.
So $D^3$ tolerates one rack failure.
\hfill \qed 

\noindent \textbf{Proof of Theorem 4.}
For each block of the $(k, l, g)$-LRC stripe, it has independent column of OA$(n,N_g^{lrc}$)
to address the nodes, so as to a region, based on the properties of OA,
traversing the all rows of OA$(n,N_g^{lrc}$), every node of the $N_g^{lrc}$ racks has the same amount blocks.
For multiple regions, OA$(r,N_g+1)$ provides the location for corresponding rack.
OA$(r,N_g+1)$ also has the same number of ids for different-type region-groups,
so the nodes among multiple racks have the same amount blocks.
In summary, data blocks, local parity blocks and global parity blocks uniformly distribute in nodes among racks.
\hfill \qed 

\subsection*{Proofs for Results in Section 5}

\noindent \textbf{Proof of Lemma 4.}
With the analysis in Section 5.1, it is easy to obtain Eq. (1). In the following, we will show that (1) is the lower bound of the average number of cross-rack accessed blocks for recovering a failed block in any data layout of distributed storage systems. So the lemma will be validated.

Upon a block failure, the recovered block should be placed in a rack such that the storage system
still tolerates a single rack failure.
So we should place the recovered block in a rack containing at most $m-1$ blocks of the same stripe before recovering.
Furthermore, since a rack contains at most $m$ blocks to tolerate a single rack failure,
an aggregated block can be computed from at most $m$ blocks in a rack.
Given a block placement, let $x$ be the number of cross-rack accessed blocks for recovering one failed block $B$.
Because we need $k$ blocks to recover the failed block, and there are at most $m-1$ blocks in the
same rack with the recovered block of $B$ and an aggregated block is constructed from at most $m$ blocks,
we have $m-1 + mx \geq k$,
i.e., $x \geq a-1 - (m-1-b)/m$.
In the of case $0 \leq b \leq m-1$,
we have that $x \geq a-1$.
Thus, $\mu$ reaches this lower bound when $b \neq  m-1$.

In case of $b=m-1$, we prove that there are at least $m-1$ blocks, where recovering each of them needs to read $x \geq a$ blocks across rack.

\begin{itemize}

    \item[(1)]
    If there is no group containing $m$ blocks, then each aggregated block are computed from at most $m-1$ blocks. Because we need at least $k$ original blocks to recover a failed block,
    $m-1 + x(m-1) \geq k$. From $k = a m-1$ in case of $b=m-1$, based on integral results, we have $x \geq a$.

    \item[(2)]
    Otherwise, there is at least one group containing $m$ blocks for two cases.

    \quad (2.1) If there are less than $a-1$ other groups containing $m$ blocks,
    then for each block in the group containing $m$ blocks, we have $x \geq a$ similar to case (1).
    So there are at least $m$ blocks with $x \geq a$.

    \quad (2.2) Otherwise, there are $a$ groups containing $m$ blocks.
    So there are $m-1$ blocks,
    where any of them does not belong to a group containing $m$ blocks.
    When any of these $m-1$ blocks fails, there are at most $m-2$ surviving blocks of the same stripe in the same rack of the failed block. Because we need $k = a  m -1$ blocks to recover a failed block and each aggregated block can be computed from at most $m$ blocks, we have $x  m + (m-2) \geq a  m -1$. So $x \geq a$.         Thus, for each of these $m-1$ blocks, we have $x \geq a$,
    i.e., there exist $m-1$ blocks with $x \geq a$.

\end{itemize}

From above, when $b=m-1$, there are $k+m = a  m +(m-1)$ blocks in a stripe, where at least $m-1$ blocks are with $x \geq a$.
With the data layout in a stripe of $D^3$,
there are $am=k+1$ blocks, which are in the groups with each group containing $m$ blocks,
satisfying that $x=a-1$,
and there are $m-1$ blocks, which are in a group containing $m-1$ blocks,
satisfying that $x=a$.
Thus, $\mu$ is exact the lower bound when $b =  m-1$.
Therefore, the lemma holds.
\hfill \qed

\noindent\textbf{Proof of Lemma 5.}
Suppose that a node in rack $R_f$ ($0 \leq f \leq N_g-1$) fails.
Within a stripe region, because the blocks in all of $n^2$ stripes are divided into groups in the same way, we can assume that $z$ blocks in each stripe of a given stripe region are stored in $R_f$.
Since there are $n^2$ stripes in a stripe region,
$R_f$ stores $z \times n^2$ blocks in the given stripe region.
From Lemma \ref{lemma:node-level},
each node in $R_f$ contains $z \times n$ blocks.
For the recovered blocks placing to a new rack,
each node in the new rack contains $z$ blocks, since $D^3$ places recovered blocks to nodes in a new rack in the round-robin order.
Therefore, the write load on the nodes in a new rack is balanced.

Within a stripe region, also because the blocks in all of $n^2$ stripes are divided into groups in the same way, we can make the assumption that the $x$-th blocks ($0 \leq x \leq len-1$) of all stripes are divided into the group with the same order, say the $k_x$-th block in the $j_x$-th group.
Let ${\cal A}_n = (a_{ij})_{ n^2\times N_g}$. Then the $x$-th block $B_{i,x}$ of $S_i$ ($0 \leq i \leq n^2-1$) is placed to $N_{j_x,((a_{ij_x}+k_x) \bmod n)}$ in rack $R_{j_x}$.
Suppose that $N_{f,f'}$ in rack $R_f$ fails ($0 \leq f \leq N_g-1, 0 \leq f' \leq n-1$)
and $z$ blocks of a stripe are stored in $R_f$.
Then the blocks $B_{i, x}$'s with $a_{if} = (f' + n - k_x) \bmod n$ and $j_x = f$ for $0 \leq i \leq n^2-1$ and $0 \leq k_x \leq z-1$ are on the failed node.
Suppose $B_{i, x'}$ is the $k_{x'}$-th block in the $j_{x'}$-th group of $S_i$,
and it participates in the recovery process for the failed block $B_{i,x}$, which means $B_{i, x'}$ may be the recovered block written into the original rack,
or the block on a node to perform inner-rack block aggregation.
According to the recovery process within a stripe, $j_{x'} \neq j_x$.

Given $k_x$ with $0 \leq k_x \leq z-1$ and $j_x$ with $j_x = f$, blocks $B_{i,x}$'s with
$a_{if} = (f' + n - k_x) \bmod n$ for $0 \leq i \leq n^2-1$
are on the failed node.
Now we are to prove that the corresponding blocks $B_{i, x'}$'s
with $a_{if} = (f' + n - k_x) \bmod n$ for $0 \leq i \leq n^2-1$ are
evenly distributed among nodes in rack $R_{j_{x'}}$.
By Definition \ref{def:oa},
in columns $j_x$ and $j_{x'}$ of ${\cal A}_n$,
every ordered pairs
$((f' + n - k_x) \bmod n, y)$
for $0 \leq y \leq n-1$ occurs in exactly one row.
So all blocks $B_{i, x'}$'s with  $a_{if} = (f' + n - k_x) \bmod n$ for $0 \leq i \leq n^2-1$
are evenly placed to nodes $N_{j_{x'},((y + k_{x'}) \bmod n)}$'s for $0 \leq y \leq n-1$ in rack $R_{j_{x'}}$,
i.e., they are evenly distributed among nodes in $R_{j_{x'}}$.
So the load are balanced among all nodes in a surviving rack in terms of overheads of read, write, and computing.
\hfill \qed 

\noindent \textbf{Proof of Theorem 5.}
Because the numbers of cross-rack accessed blocks
for recovering a block in all stripes are minimized. So from
Lemma \ref{lemma:recovery-stripe-level}, the theorem holds.
\hfill \qed 

\noindent \textbf{Proof of Theorem 6.}
Let ${\cal M} = (m_{ij})_{(r(r-1))\times (N_g+1)}$.
Then region-group $G^i_j$
is placed in rack $R_{m_{ij}}$ for $0 \leq i \leq r(r-1)-1, 0 \leq j \leq N_g-1$.
Suppose that a node in rack $R_f$ ($0 \leq f \leq r-1$) fails.
Then the $G^i_j$'s with $m_{ij} = f$
have blocks on the failed node.
Thus, given $x$ for $0 \leq x \leq N_g-1$,
the $G^i_x$'s with $m_{ix} = f$
have blocks on the failed node.
Suppose that $G^i_{x'}$ and $H_i$ (if any) participate
in the recovery for the failed blocks in $G^i_x$.
According to the recovery process within a stripe region,
we have $x' \neq x$.
By Definition \ref{def:oa} and the definition of ${\cal M}$,
in columns $x$ and $y$ of ${\cal M}$,
where $y=x'$ for $G^i_{x'}$ or $y=N_g$ for $H_i$,
every ordered pairs $(f,j)$ for $0 \leq j \leq r-1$ and $j \neq f$ occurs in exactly one row.
Then the $G^i_{x'}$'s and the $H_i$'s are
in the racks $R_j$'s for $0 \leq j \leq r-1$ and $j \neq f$.
This implies that for each $x$ ($0 \leq x \leq N_g-1$),
the $G^i_{x'}$'s and the $H_i$'s with $m_{ix} = f$
for $0 \leq i \leq r(r-1)-1$
are evenly distributed among surviving racks.
Therefore, the cross-rack read/write load is balanced among surviving racks.

Furthermore, by Lemma \ref{lemma:recovery-node-level},
the loads for recovering a single node failure are balanced among nodes in each surviving rack within a stripe region in terms of overheads of read, write and computing respectively.
Therefore, the loads are balanced among all nodes in surviving racks in terms of read, write and computing respectively.

From all above, the theorem holds.
\hfill \qed 

\noindent \textbf{Proof of Theorem 7.}
From Theorem \ref{theorem:lrc-data-layout}, the data blocks, local parity blocks
and global parity blocks are uniformly distributed across nodes,  respectively.
When recovering single-node failure,
according to the distribution of $D^3$,  all blocks in a local group are  corresponding to different columns of ${\cal A}$, so it reaches repair traffic balance across nodes in a rack for recovering a failed data/local parity block by Property \ref{def:pro2} of OA. Similarly, it also reaches balanced repair traffic for recovering a failed global parity block. 
Load balance across racks of $D^3$ further guarantees repair traffic balance across all surviving nodes, which means load balance of reading data.
For recovered blocks, the decoded data will be written to replacement nodes, chosen in a round-robin way in the rack identified by the last column of ${\cal M}$.
Property \ref{def:pro1} of OA guarantees uniform distribution of replacement nodes.
So,  we achieve load balance of write and computing among surviving nodes.
\hfill \qed 

\noindent \textbf{Proof of Theorem 8.}
We get optimal trade off from the above discussion of minimum one-batch data traffic.
About load balance of migrated data, because we achieve the load balance of write in recovering state (see Theorem \ref{theorem:recovery-rack-level-2} and \ref{lrc_recovery}), one-batch migration will involve the same number blocks in node and rack level, and then we achieve load balance among surviving racks in per-batch migration. Since there are $(r-1)*N_g$ failed-rack ids in the first $N_g$ columns of $\cal{M}$, the DSS will rearrange racks for the failed blocks in failed racks. The failed blocks belong to one region will choose the same new racks.
Each-batch migration is involved in $r-1$ region-groups which mean $r-1$ different racks, so the DSS needs do $N_g$ batches migration.
\hfill \qed